\documentclass[hidelinks,onefignum,onetabnum]{siamart251216}
\usepackage{amssymb}
\usepackage{graphicx}
\usepackage{subfig}
\usepackage{mathrsfs}
\usepackage{booktabs}
\usepackage{multirow}
\usepackage{threeparttable}
\usepackage{placeins}
\renewcommand{\vec}[1]{\boldsymbol{#1}}
\title{A multispecies unified gas-kinetic scheme for coupled kinetic material and radiative transport}
\author{Mingyu Quan\thanks{Department of Mathematics, University of British Columbia, 1984 Mathematics Road, Vancouver, BC, Canada V6T 1Z2 (\email{mquan@math.ubc.ca}).}}
\headers{Multispecies kinetic and radiative transport}{M. Quan}
\begin{document}
\maketitle
\begin{abstract}
We develop a multispecies unified gas-kinetic scheme for coupled electron, ion, and grey-radiation transport. Building on our earlier GKS--UGKS formulation, the method evolves electron and ion distribution functions in discrete velocity space and incorporates the velocity-dependent radiation--electron energy and momentum exchange terms, including contributions from the radiative flux and pressure tensor. The material fluxes retain species-dependent departures from local equilibrium, while angularly resolved radiation transport describes the radiative field. Electrons exchange momentum and energy directly with radiation and with ions through interspecies collisions. Both material and radiation interface fluxes are constructed from time-dependent kinetic solutions. Binary-mixture shock and Marshak-wave benchmarks assess the material and radiation components, respectively. A three-temperature shock tube, radiative shocks, and a heterogeneous two-dimensional Tophat problem then demonstrate coupled wave structures and distinct electron, ion, and radiation responses. The resulting framework combines species-resolved kinetic transport with velocity-dependent radiation--electron coupling for radiation--plasma flows across different material and radiative transport conditions.
\end{abstract}
\section{Introduction}\label{sec:intro}
Radiation--plasma flows involve several forms of nonequilibrium. Radiation can have an anisotropic angular distribution, electrons and ions can have different velocities and temperatures, and each material distribution can depart from its local Maxwellian. Photon interactions and material collisions control these processes on different scales. For example, an optically thick region can support radiative diffusion while its material remains close to local equilibrium, whereas a neighboring dilute region can allow both photons and material particles to travel over longer distances between interactions. Such conditions arise in high-energy-density and astrophysical flows \cite{drake2006,mihalas1999,atzeni2004,jiang2019super}. Their simulation calls for a coupled description of radiative transport, species-dependent material transport, and momentum and energy exchange.

Radiation transport has been treated by particle methods, solvers for the transfer equation, and radiation-hydrodynamic models. Three-temperature models evolve electron, ion, and radiation energies within a continuum description \cite{Enaux2022,Peng2022,cheng2024}. Solving for the radiation intensity also retains its angular dependence outside the diffusion regime \cite{lowrie1999,jiang2014algorithm}. Methods that connect free transport and diffusion must account for both photon propagation and interactions with the material \cite{tang2021front}. Radiation unified gas-kinetic schemes (UGKS) accomplish this through interface fluxes constructed from time-dependent kinetic solutions \cite{mieussens2013,sun2015gray,sun2015frequency}. Coupling these radiation solvers to a material model introduces a further choice: whether electrons and ions are represented by continuum variables or by their velocity distributions.

Our earlier GKS--UGKS formulation \cite{xu2025} used a continuum gas-kinetic solver for the electron and ion components and a UGKS for grey radiative transfer. The present work extends that formulation in two respects. First, separate microscopic electron and ion distribution functions become numerical unknowns. This permits species-dependent kinetic departures from local equilibrium to enter the material transport, extending the description to rarefied conditions. Second, the radiation--electron interaction retains velocity-dependent energy and momentum exchange contributions involving the radiation flux and pressure tensor. These contributions couple the motion of electrons to the angular structure of the radiation field.

The material and radiation fluxes are both derived from kinetic evolution over a numerical time step. For the material subsystem, transport and relaxation jointly determine the interface distribution, following the multiscale construction of UGKS \cite{xu2010,xu2022implicit}. Radiation interacts directly with electrons, and interspecies collisions transmit momentum and energy to ions. The resulting formulation therefore represents material kinetic nonequilibrium and radiation--material exchange within the same coupled update.

The numerical study progresses from component benchmarks to coupled physical configurations. Binary-mixture shocks compare species-dependent profiles with Boltzmann solutions, and a Marshak wave examines radiation diffusion and thermal exchange in stationary material. The three-temperature shock tube connects the coupled formulation to a continuum radiation-hydrodynamic comparison. Radiative shocks exhibit the interaction of compression, radiative heating, and thermal relaxation, while the two-dimensional Tophat configuration examines distinct material and radiation responses across adjacent opaque and dilute regions. These complementary cases make the contribution of the kinetic material description and the behavior of the coupled exchange processes explicit.

Section~\ref{sec:rad-ugks} specifies the radiative model and its discretization. Section~\ref{sec:fluid-ugks} presents the material UGKS and the coupled update. Section~\ref{sec:case} reports the numerical comparisons, and Section~\ref{sec:conclusion} summarizes their implications and limitations.

\section{Kinetic model and numerical algorithm for radiative transfer}\label{sec:rad-ugks}

\subsection{Kinetic model}

We describe electrons and ions using the multispecies relaxation model of Andries, Aoki, and Perthame \cite{aap2002}, and radiation using a grey transfer equation with material-velocity corrections. This multispecies kinetic model satisfies the indifferentiability principle: when the species have identical properties, it reduces to the single-species model. It also satisfies an H-theorem and reproduces the interspecies momentum and energy exchange relations for Maxwell molecules. The numerical interaction coefficients used here are specified in Section~\ref{sec:material-exchange}.

The radiation intensity is measured in the fixed spatial frame, while absorption and scattering are described relative to the moving material \cite{lowrie1999,jiang2014algorithm}. The velocity-dependent terms account for this relation between frames and for radiative momentum deposition. The radiation equation used in the coupled calculation is

{\footnotesize
\begin{equation} \label{eq:kinetic-model}
\begin{aligned}
	&\frac{\partial I}{\partial t} +c\vec{\Omega} \cdot \frac{\partial I}{\partial \vec{x}} = c\sigma_{a}\biggl(\frac{a_{\mathcal R}T^{4}}{4\pi}-I\biggr) + c\sigma_{s}(J - I) + 3\vec{\Omega}\cdot\vec{U}\sigma_{a}\biggl( \frac{a_{\mathcal R}T^{4}}{4\pi}-J\biggr) \\
	& \qquad \qquad  + \vec{\Omega}\cdot\vec{U}(\sigma_{a}+\sigma_{s})(I+3J)- 2\sigma_{s}\vec{U}\cdot\vec{H} - (\sigma_{a}-\sigma_{s})\biggl(\frac{\vec{U}\cdot\vec{U}}{c}J + \frac{\vec{U}\cdot(\vec{U}\cdot\textbf{K})}{c}\biggr) \triangleq Q_{\mathcal{R}},
\end{aligned}
\end{equation}
}

Here $I(\vec{x},\vec{\Omega},t)$ is the energy-normalized, frequency-integrated radiation intensity at position $\vec{x}$, time $t$, and propagation direction $\vec{\Omega}$, a unit vector. The grey approximation integrates over photon frequency while retaining this angular dependence. The coefficients $\sigma_a$ and $\sigma_s$ describe absorption and scattering, respectively; $a_{\mathcal R}$ is the radiation constant and $c$ the speed of light. In the radiation--electron update, $T$ and $\vec{U}$ are the electron temperature and bulk velocity. We use ordinary solid angle on the full sphere, with $\int \mathrm{d}\vec{\Omega}=4\pi$. The angular averages appearing in the velocity corrections are defined by

\begin{displaymath}
\begin{aligned}
	&{J} \equiv \frac{1}{4\pi}\int I {\rm d} \vec{\Omega},\\
	&\vec{H} \equiv \frac{1}{4\pi}\int \vec{\Omega}I {\rm d} \vec{\Omega},\\
	&\textbf{K} \equiv \frac{1}{4\pi}\int \vec{\Omega} \otimes \vec{\Omega}I {\rm d} \vec{\Omega}.
\end{aligned}
\end{displaymath}
The scalar $J$ is the angular mean intensity. The vector $\vec{H}$ measures the directional imbalance of radiation, and the tensor $\textbf{K}$ describes its angular second moment. These quantities provide the angular information entering the energy and momentum exchange; they are evaluated from the intensity rather than prescribed by a diffusion closure. The dyadic product $\vec{\Omega}\otimes\vec{\Omega}$ is a second-order tensor.

For flux construction, we collect the velocity-dependent terms in Eq.~\eqref{eq:kinetic-model} into $S$. Thus $S$ is the velocity correction, while $Q_{\mathcal{R}}$ denotes the entire angular source, including emission, absorption, and scattering:

\begin{equation}
\begin{aligned}
	S = &3\vec{\Omega}\cdot\vec{U}\sigma_{a}\biggl( \frac{a_{\mathcal R}T^{4}}{4\pi}-J\biggr) + \vec{\Omega}\cdot\vec{U}(\sigma_{a}+\sigma_{s})(I+3J)\\
	      &- 2\sigma_{s}\vec{U}\cdot\vec{H} - (\sigma_{a}-\sigma_{s})\biggl(\frac{\vec{U}\cdot\vec{U}}{c}J + \frac{\vec{U}\cdot(\vec{U}\cdot\textbf{K})}{c}\biggr) .
\end{aligned}
\end{equation}

Angular integration gives the macroscopic energy source $S_{\mathcal{R}}(E)$ and momentum source $\vec{S}_{\mathcal{R}}(\vec{P})$ \cite{lowrie1999,jiang2014algorithm}. The arguments $(E)$ and $(\vec{P})$ identify the exchanged conserved quantity and distinguish these moment sources from the angular source $Q_{\mathcal{R}}$. The coupling retains the material-velocity, radiative-flux, and radiation-pressure contributions that were simplified in our earlier formulation \cite{xu2025}:
\begin{equation}{\label{eq:source}}
\begin{aligned}
	&S_{\mathcal{R}}(E) =\sigma_a(a_{\mathcal{R}}T^4-E_{\mathcal{R}}) +(\sigma_a-\sigma_s)\frac {\vec{U}}{c^2}\cdot[\vec{F}_{\mathcal{R}}-(\vec{U}E_{\mathcal{R}}+\vec{U}\cdot\textbf{P}_{\mathcal{R}})], \\
	&\vec{S}_{\mathcal{R}}(\textbf{P}) =-\frac{(\sigma_s+\sigma_a)}c[\vec{F}_{\mathcal{R}}-(\vec{U}E_{\mathcal{R}}+\vec{U}\cdot\textbf{P}_{\mathcal{R}})] +\frac {\vec{U}}{c}\sigma_a(a_{\mathcal{R}}T^4-E_{\mathcal{R}}),
\end{aligned}
\end{equation}
where the radiation moments are
\begin{equation}\label{eq:radiation-moments}
\begin{aligned}
E_{\mathcal R}&=\int I\,\mathrm{d}\vec\Omega=4\pi J,\\
\vec F_{\mathcal R}&=c\int\vec\Omega I\,\mathrm{d}\vec\Omega=4\pi c\vec H,\\
\textbf P_{\mathcal R}&=\int\vec\Omega\otimes\vec\Omega I\,\mathrm{d}\vec\Omega=4\pi\textbf K.
\end{aligned}
\end{equation}
Thus $E_{\mathcal R}$, $\vec F_{\mathcal R}$, and $\textbf P_{\mathcal R}$ are radiation energy density, energy flux, and pressure tensor, respectively. With this intensity convention, the angular source satisfies
\begin{equation}\label{eq:source-moments}
\int Q_{\mathcal R}\,\mathrm{d}\vec\Omega=cS_{\mathcal R}(E),\qquad
\frac1c\int\vec\Omega Q_{\mathcal R}\,\mathrm{d}\vec\Omega=\vec S_{\mathcal R}(\vec P).
\end{equation}
These are radiation-side sources; the corresponding material energy and momentum sources are $-cS_{\mathcal R}(E)$ and $-\vec S_{\mathcal R}(\vec P)$. The term proportional to $a_{\mathcal{R}}T^4-E_{\mathcal{R}}$ represents emission and absorption driven by thermal disequilibrium. The flux and pressure terms account for the relative motion of radiation and material and couple radiative momentum transfer to energy exchange.

\subsection{Numerical Discretization and Flux Construction}

Let $V_i$ be the volume of cell $i$, $N(i)$ its set of neighboring cells, $\mathcal{A}_{ij}$ the area of the shared face, and $\vec{n}_{ij}$ its outward unit normal from cell $i$. Superscripts $n$ and $n+1$ denote the beginning and end of a step of length $\Delta t$. The finite volume update for each discrete radiation direction is
\begin{equation}\label{eq:I}
	I_i^{n+1}=I_i^n-\frac{\Delta t}{V_i} \sum_{j \in N(i)} \mathcal{F}_{i j, \mathcal{R}} \mathcal{A}_{i j}+\int_0^{\Delta t}Q_{\mathcal{R}} {\rm d} t,
\end{equation}
The last term integrates the local angular source over the time step. The time-averaged intensity flux $\mathcal{F}_{ij,\mathcal{R}}$ through face $ij$ is
\begin{equation}
	\mathcal{F}_{ij,\mathcal{R}}=\frac{c}{\Delta t}\int_{0}^{\Delta t} \vec{\Omega}\cdot \vec{n}_{ij} I_{ij}(t) {\rm d}t,
\end{equation}
where $I_{ij}(t)$ is the evolving interface intensity. Its characteristic solution contains the transported initial intensity and the contributions generated by emission, scattering, and the velocity correction:
\begin{equation}\label{eq:rad-characteristic}
\begin{aligned}
I(\vec x_0,t)={}&e^{-\nu t}I^0(\vec x_0-c\vec\Omega t)\\
&+\int_0^t e^{-\nu(t-t')}
\left[\frac{c\sigma_a}{4\pi}\phi+c\sigma_s\psi+S\right]
(\vec x_0-c\vec\Omega(t-t'),t')\,\mathrm{d}t'.
\end{aligned}
\end{equation}
Here $\vec x_0$ is the interface location, $t'$ an integration time within the current step, and $I^0$ the initial intensity. We set $\phi=a_{\mathcal R}T^4$, $\psi=J$, and $\nu=c(\sigma_a+\sigma_s)$. The absorption and scattering coefficients are held locally fixed during flux integration. The exponential therefore accounts for the full extinction in Eq.~\eqref{eq:kinetic-model}. Using $\vec\xi=\vec x-\vec x_0$, the interface reconstructions are
\begin{equation}
\begin{aligned}
\phi(\vec x,t)&=\phi^0+\nabla\phi\cdot\vec\xi+\partial_t\phi\,t+\mathcal O(|\vec\xi|^2+t^2),\\
\psi(\vec x,t)&=\psi^0+\nabla\psi\cdot\vec\xi+\partial_t\psi\,t+\mathcal O(|\vec\xi|^2+t^2),\\
I^0(\vec x)&=I^0+\nabla I^0\cdot\vec\xi+\mathcal O(|\vec\xi|^2).
\end{aligned}
\end{equation}
The derivatives are evaluated at the interface at the beginning of the step. For the flux expression below, the velocity correction $S$ is approximated by its local interface value during the integration.

Substituting the reconstructions into the characteristic solution and averaging over the step gives the directional radiation flux
\begin{equation}\label{eq:micro-flux-rad}
	\begin{aligned}
		\mathcal{F}_{ij,\mathcal{R}}
		=
		& \vec{\Omega}\cdot \vec{n}_{ij}\left(
		A_1 I^0 + A_2 \frac{\partial I}{\partial \vec{x}} \cdot \vec{\Omega}
		\right)
		+ \vec{\Omega}\cdot \vec{n}_{ij} \left(
		A_3 \phi^{0}
		+ A_4 \frac{\partial \phi}{\partial \vec{x}}\cdot \vec{\Omega}
		+ A_5 \frac{\partial \phi}{\partial t}
		\right)\\
		&+ \vec{\Omega}\cdot \vec{n}_{ij} \left(
		A_6 \psi^{0}
		+ A_7 \frac{\partial \psi}{\partial \vec{x}}\cdot \vec{\Omega}
		+ A_8 \frac{\partial \psi}{\partial t}\right)
		+\vec{\Omega}\cdot \vec{n}_{ij} \left(  A_9 S\right)
	\end{aligned}
\end{equation}
The groups with $A_1$ and $A_2$ arise from the initial intensity; $A_3$--$A_5$ describe thermal emission, $A_6$--$A_8$ scattering, and $A_9$ the velocity correction. Their time-integration coefficients are
\begin{equation}\label{eq:rad-coefficients}
\begin{aligned}
A_1&=\frac{c}{\Delta t\nu}(1-e^{-z}),\\
A_2&=-\frac{c^2}{\Delta t\nu^2}(1-e^{-z}-ze^{-z}),\\
A_3&=\frac{c^2\sigma_a}{4\pi\Delta t\nu}\left[\Delta t-\frac{1-e^{-z}}\nu\right],\\
A_4&=-\frac{c^3\sigma_a}{4\pi\Delta t\nu^2}\left[\Delta t(1+e^{-z})-\frac{2(1-e^{-z})}\nu\right],\\
A_5&=\frac{c^2\sigma_a}{4\pi\Delta t\nu^3}\left[\frac{z^2}{2}-z+1-e^{-z}\right],\\
A_6&=\frac{c^2\sigma_s}{\Delta t\nu}\left[\Delta t-\frac{1-e^{-z}}\nu\right],\\
A_7&=-\frac{c^3\sigma_s}{\Delta t\nu^2}\left[\Delta t(1+e^{-z})-\frac{2(1-e^{-z})}\nu\right],\\
A_8&=\frac{c^2\sigma_s}{\Delta t\nu^3}\left[\frac{z^2}{2}-z+1-e^{-z}\right],\\
A_9&=\frac{c}{\Delta t\nu}\left[\Delta t-\frac{1-e^{-z}}\nu\right],\qquad z=\nu\Delta t.
\end{aligned}
\end{equation}
The coefficients follow by integrating Eq.~\eqref{eq:rad-characteristic} over the time step. Their removable singularities at $\nu=0$ are interpreted by continuity. With vanishing opacities, the source contributions vanish and $A_1\to c$, $A_2\to-c^2\Delta t/2$, recovering the reconstructed free-transport flux.

For the three-temperature kinetic model used in this paper, the radiation energy $E_{\mathcal{R}}$ is obtained by taking moment of the radiation intensity. The discretized macroscopic equation can be derived by taking moments of Eq.~\eqref{eq:I}
\begin{equation}
	(E_{\mathcal{R}})_i^{n+1}=(E_{\mathcal{R}})_i^n-\frac{\Delta t}{V_i} \sum_{j \in N(i)} {F}_{ij,\mathcal{R}} \mathcal{A}_{i j} + c\int_{0}^{\Delta t} S_{\mathcal{R}}(E) {\rm d} t,
\end{equation}
where $F_{ij, \mathcal{R}}$ denotes the macroscopic flux for radiation energy at the cell interface $ij$

\begin{equation}\label{eq:macro-flux-rad}
	F_{ij,\mathcal{R}} = \int \mathcal{F}_{ij,\mathcal{R}} {\rm d} \vec{\Omega} = \sum_k w_k\mathcal{F}_{ij,\mathcal R}(\vec\Omega_k),
\end{equation}
The directions $\vec\Omega_k$ and weights $w_k$ define the angular quadrature, with $\sum_k w_k=4\pi$. The scalar face flux $F_{ij,\mathcal R}$ is distinct from the physical vector energy flux $\vec F_{\mathcal R}$.

\section{Governing equations and numerical algorithm for fluid flow}\label{sec:fluid-ugks}

\subsection{Governing equations and kinetic model}

The material update combines species transport, electron--ion collisions, and radiation--electron exchange. The following balance laws summarize the mass, momentum, and total-energy variables used in this update \cite{jiang2014global}: 
\begin{equation}\label{eq:mhd}
	\begin{aligned}
		&\partial_{t}\rho+\partial_{\vec{x}}\cdot(\rho\vec{U})=0,\\
		&\partial_{t}(\rho\vec{U})+\partial_{\vec{x}}\cdot(\rho\vec{U}\vec{U}+p \vec{\rm{D}} )= -\vec{S}_{\mathcal{R}}(\vec{P}),\\
		&\partial_{t}E+\partial_{\vec{x}}\cdot\left((E+p)\vec{U}\right)= -cS_{\mathcal{R}}(E),
	\end{aligned}
\end{equation}
Here $\rho$, $\vec{U}$, $p$, and $E$ are the material mass density, bulk velocity, pressure, and total-energy density, and $\vec{\rm D}$ is the identity tensor. The radiation-side sources are given in Eq.~\eqref{eq:source}; their negatives, with the energy factor $c$, act on electrons in the species update below. This pairing conserves total material--radiation energy and momentum during local exchange. The microscopic material description is
\begin{equation}\label{eq:kinetic-1}
	\frac{\partial f_{\alpha}}{\partial t} + \vec{u}_{\alpha}\cdot\frac{\partial f_{\alpha}}{\partial \vec{x}} = \frac{g_{\alpha}^{M} - f_{\alpha}}{\tau_{\alpha}}, \alpha \in \mathcal{I}, \mathcal{E}, 
\end{equation}
The index $\alpha=\mathcal{E},\mathcal{I}$ identifies electrons and ions. The function $f_\alpha(\vec{x},\vec{u},t)$ is their mass-weighted velocity distribution, where $\vec{u}$ is microscopic velocity and $\vec{U}_\alpha$ its bulk mean. The relaxation time is $\tau_\alpha$. The target $g_\alpha^M$ is the multispecies Maxwellian of the AAP model: its moments account for exchange with the other species. The split implementation first uses the species Maxwellian in Eq.~\eqref{eq:eq-state} for transport, then applies the exchange relations in Eq.~\eqref{eq:aap-exchange}.

\subsection{Numerical algorithm}
Within the finite volume framework, the update of the macroscopic variables $\vec{W}$ in $V_i$ at a discrete time scale $\Delta t = t^{n+1}-t^{n}$ satisfies the fundamental conservation laws
\begin{equation}\label{eq:fvm-f}
	\vec{W}_{i}^{n + 1}=\vec{W}_{i}^n - \frac{\Delta t}{V_i} \sum\limits_{j \in N(i)} {\vec{F}}_{ij} {\mathcal A}_{ij} +\vec{S}_{i},
\end{equation}
where $V_{i}$ is the volume of cell $i$, $\vec{W}_{i}$ is the set of conservative variables of the cell $i$, i.e., the densities of mass $\rho$, momentum $\rho{\vec U} $, and total energy density $E$.  $N(i)$ denote the set of neighboring cells that share a common interface with cell $i$. For each $j \in N(i)$, the subscript $ij$ refers to quantities defined on the interface between cells $i$ and $j$. Hence, ${\mathcal A}_{ij}$ is referred to as the area of the interface $ij$.

The material evolution is split into three stages. Stage $P1$ transports each species with its UGKS interface flux. Stage $P2$ applies collisional momentum and energy exchange between electrons and ions. Stage $P3$ applies radiation--electron exchange, while the ion state is held fixed during that stage. Thus radiation affects ions through the subsequent interspecies dynamics.

\begin{equation}\label{eq:step}
\begin{aligned}
	&P1 =
	\left\{ \begin{array} {l l l}
		{\partial_{t}\rho_{\alpha}+\partial_{\vec{x}}\cdot(\rho_{\alpha}\vec{U}_{\alpha})=0,}\\
		{\partial_{t}(\rho_{\alpha}\vec{U}_{\alpha})+\partial_{\vec{x}}\cdot(\rho_{\alpha}\vec{U}_{\alpha}\vec{U}_{\alpha}+p_{\alpha} \vec{\rm{D}})=  0,}\\
		{\partial_{t}E_{\alpha}+\partial_{\vec{x}}\cdot\left((E_{\alpha}+p_{\alpha})\vec{U}_{\alpha}\right)= 0,}\end{array} \right.\\
	&P2 =
	\left\{ \begin{array} {l l l}
		{\partial_{t}\rho_{\alpha}=0,}\\
		{\partial_{t}(\rho_{\alpha}\vec{U}_{\alpha})= \vec{S}_{\alpha},}\\
		{\partial_{t}E_{\alpha} = S_{\alpha},}\end{array} \right.\\
	&P3 =
	\left\{ \begin{array} {l l l l l l}
		{\partial_{t}\rho_{\mathcal{E}}=0,}\\
		{\partial_{t}(\rho_{\mathcal{E}}\vec{U}_{\mathcal{E}})= -\vec{S}_{\mathcal{R}}(\vec{P}),}\\
		{\partial_{t}E_{\mathcal{E}} = -cS_{\mathcal{R}}(E),}\\
		{\partial_{t}\rho_{\mathcal{I}}=0,}\\
		{\partial_{t}(\rho_{\mathcal{I}}\vec{U}_{\mathcal{I}})= 0,}\\
		{\partial_{t}E_{\mathcal{I}} = 0,}\end{array} \right.
\end{aligned}
\end{equation}
The superscripts $*$ and $**$ identify the states after material transport and interspecies exchange, respectively:
\begin{equation}
	P1:\mathbf{W}^n\to\mathbf{W}^*,\quad P2:\mathbf{W}^*\to\mathbf{W}^{**},\quad P3:\mathbf{W}^{**}\to\mathbf{W}^{n+1}.
\end{equation}

\subsubsection[Part 1: Transport update]{Part 1: $\mathbf{W}^n\to\mathbf{W}^*$}
For the construction of the numerical integral solution in the UGKS framework, we restate the governing kinetic equation in a form suitable for characteristic integration. The kinetic model for $P1$ is
\begin{equation}\label{eq:kinetic-fluid}
 	\frac{\partial f_{\alpha}}{\partial t} + \vec{u}_{\alpha}\cdot\frac{\partial f_{\alpha}}{\partial \vec{x}} = \frac{g_{\alpha} - f_{\alpha}}{\tau_{\alpha}},
\end{equation}
For this transport stage, $g_\alpha$ is the Maxwellian with the density, velocity, and temperature of species $\alpha$: 
\begin{equation}\label{eq:eq-state}
g_{\alpha} =\rho_{\alpha} \left(\frac{m_{\alpha} }{2\pi k_BT_{\alpha}}\right)^{3/2}\exp\left(-\frac{m_{\alpha} }{2k_{B}T_{\alpha}}\left(\vec{u}-\vec{U}_{\alpha} \right)^{2}\right),
\end{equation}
Here $m_\alpha$ is the species particle mass and $k_B$ the Boltzmann constant. The parameters $\rho_\alpha$, $\vec{U}_\alpha$, and $T_\alpha$ are obtained from moments of $f_\alpha$. Relaxation toward this local Maxwellian preserves species mass, momentum, and energy during $P1$, as expressed by
\begin{equation}
	\int \frac{g_{\alpha} - f_{\alpha}}{\tau_{\alpha}} \vec{\psi} {\rm d} \vec{u} = 0,
\end{equation}
with $\vec{\psi} = (1, \vec{u}, \frac12 \vec{u}^2)^T$. The time-dependent distribution function $f_{ij}(t)$  on the interface is constructed by the integral solution of the kinetic model Eq.~\eqref{eq:kinetic-fluid}
\begin{equation}
	f(\vec{x},t)={\frac{1}{\tau}}\int_{0}^{t}g(\vec{x}^{\prime},t^{\prime})e^{-(t-t^{\prime})/\tau}{\rm d} t^{\prime}+e^{-t/\tau}f_{0} (\vec{x}-\vec{u}t),
\end{equation}
Here $f_0$ is the distribution at the beginning of the step and $\vec{x}'=\vec{x}-\vec{u}(t-t')$ follows the particle characteristic. The integral term accumulates relaxation toward equilibrium, while the exponential term carries the initial distribution along free transport. Their relative weights depend on $\Delta t/\tau_\alpha$, retaining the contribution of the nonequilibrium distribution to the interface flux. Linear reconstruction supplies the spatial and temporal terms:
\begin{equation}\label{eq:Taylor}
	\begin{aligned}
		&g(\vec{x},t) = g^0
		+ \vec{x} \cdot \frac{\partial g }{\partial \vec{x}}
		+ \frac{\partial g }{\partial t} t, \\
		&f^0_{\alpha}(\vec{x})=f_{\alpha}^{l,r}
		+\vec{x} \cdot \frac{\partial f_{\alpha}^{l,r}}{\partial \vec{x}},
		\end{aligned}
	\end{equation}
where $f_{\alpha}^{l,r}$ is the distribution function constructed by distribution functions $f_\alpha^l$ and $f_\alpha^r$ interpolated from cell centers to the left and right sides of the interface
\begin{equation*}
f_{\alpha}^{l,r} = f_{\alpha}^l H\left[\bar{u}_{ij}\right]
+ f_\alpha^r\left(1-H\left[\bar{u}_{ij}\right]\right),
\end{equation*}
where $\bar{u}_{ij}=\vec{u}\cdot\vec{n}_{ij}$ is the particle velocity projected on the normal direction of the cell interface $\vec{n}_{ij}$, and $H[x]$ is the Heaviside function. The equilibrium state $g^0_{\alpha}$ at the cell interface is computed from the colliding particles from both sides of the cell interface,
\begin{equation}
\vec{W}^0_\alpha
= \int g^0_\alpha \vec{\psi} {\rm d} \vec{u}
= \int f^{l,r}_\alpha \vec{\psi} {\rm d} \vec{u},
\end{equation}
and the spatial and temporal derivatives of the equilibrium state can be obtained by
\begin{equation}
\begin{aligned}
& \int \frac{\partial g_\alpha}{\partial \vec{x}} \vec{\psi}  {\rm d}  \vec{u}
= \frac{\partial \vec{W}_\alpha}{\partial \vec{x}}, \\
& \int \frac{\partial g_\alpha}{\partial t} \vec{\psi}  {\rm d}  \vec{u}
= - \int \vec{u} \cdot \frac{\partial g_\alpha}{\partial \vec{x}} \vec{\psi}  {\rm d}  \vec{u},
\end{aligned}
\end{equation}
where the spatial derivatives of the conservative variables ${\partial \vec{W}_\alpha}/{\partial \vec{x}}$ can be obtained from reconstruction. 

The time-averaged  microscopic flux ${\mathcal F}_{ij}$  and macroscopic flux $\vec{F}_{ij}$ can be expressed as
\begin{equation}\label{eq:micro-flux}
	\mathcal{F}_{ij} = \frac{1}{\Delta t} \int_0^{\Delta t} \vec{u} \cdot \vec{n}_{ij} f_{ij}(t) {\rm d}t,
\end{equation}
and
\begin{equation}\label{eq:macro-flux-aap}
	\vec{F}_{ij} = \int \mathcal{F}_{ij} \vec{\psi} {\rm d} \vec{u} = \sum_k \omega_k\mathcal{F}_{ij,k}\vec{\psi}_{k},
\end{equation}
The index $k$ labels discrete material velocities $\vec u_k$, and $\omega_k$ is the corresponding velocity-space quadrature weight. On a uniform Cartesian midpoint grid, $\omega_k$ is the volume of a velocity-space cell. Microscopic and macroscopic fluxes are obtained from the same evolving interface distribution.
Substituting Eq~\eqref{eq:Taylor} into Eq~\eqref{eq:micro-flux}, we can get the time-averaged microscopic flux
\begin{equation}\label{eq:micro-flux-aap}
\mathcal{F}_{ij,\alpha}=\vec{u} \cdot \vec{n}_{ij}
\left(
C_{1}g^0_{\alpha}+C_{2} \frac{\partial g_\alpha}{\partial \vec{x}} \cdot \vec{u}+C_{3} \frac{\partial g_\alpha}{\partial t}
\right)
+ \vec{u} \cdot \vec{n}_{ij}
\left(
C_4 f_\alpha^{l,r}
+ C_5 \vec{u} \cdot \frac{\partial f_\alpha^{l,r}}{\partial \vec{x}},
\right)
\end{equation}
The $C_1$--$C_3$ terms describe equilibrium relaxation and its reconstructed variation, while $C_4$ and $C_5$ retain the freely transported initial distribution. The coefficients are \cite{xu2010,xu2014,sun2015gray}
\begin{equation*}
\begin{aligned}
C_1 &= 1 - \frac{\tau}{\Delta t} \left( 1 - e^{-\Delta t / \tau} \right) , \\
C_2 &= -\tau + \frac{2\tau^2}{\Delta t} - e^{-\Delta t / \tau} \left( \frac{2\tau^2}{\Delta t} + \tau\right) ,\\
C_3 &=  \frac12 \Delta t - \tau + \frac{\tau^2}{\Delta t} \left( 1 - e^{-\Delta t / \tau} \right) , \\
C_4 &= \frac{\tau}{\Delta t} \left(1 - e^{-\Delta t / \tau}\right), \\
C_5 &= \tau  e^{-\Delta t / \tau} - \frac{\tau^2}{\Delta t}(1 -  e^{-\Delta t / \tau}).
\end{aligned}
\end{equation*}

The intermediate states are updated by 
\begin{equation}\label{eq:update-mid-1}
	\vec{W}_{i}^{*} = {\vec W}_{i}^{n}-\frac{\Delta t}{V_{i}} \sum\limits_{j \in N(i)}\vec{F}_{ij} {\mathcal{A}}_{ij}.
\end{equation}

\subsubsection[Part 2: Interspecies exchange]{Part 2: $\mathbf{W}^*\to\mathbf{W}^{**}$}\label{sec:material-exchange}
In the second stage, the fluid variables are updated through momentum and energy exchange between electrons and ions.
The macroscopic velocity $\vec{U}_\alpha^{* *}$ and temperature $T_\alpha^{* *}$ are determined after momentum and energy exchange by multispecies collisions between electrons and ions \cite{xiao2019unified,aap2002},

\begin{equation}\label{eq:aap-exchange}
	\begin{aligned}
		&\vec{U}_{\alpha}^{**} = \vec{U}_{\alpha} ^{*}+ \tau_{\alpha}\sum\limits^N\limits_{k = \mathcal{E}}\frac{2\rho_{k} ^{*}\chi_{\alpha k}}{m_{\alpha} + m_k}(\vec{U}_{k} ^{*} - \vec{U}_{\alpha} ^{*}), \\
		&T_{\alpha}^{* *} = T_{\alpha} ^{*} - \frac{m_{\alpha}}{3k_{B}}(\vec{U}_{\alpha}^{* *} - \vec{U}_{\alpha} ^{*})^2  + \tau_{\alpha}\sum\limits^N\limits_{k = \mathcal{E}}\frac{4\rho_{k} ^{*}\chi_{\alpha k}}{(m_{\alpha} + m_k)^2}\left(T_{k} ^{*} - T_{\alpha}  ^{*}+ \frac{m_{k}}{3k_{B}}(\vec{U}_{k} ^{*} - \vec{U}_{\alpha} ^{*})^2\right),
	\end{aligned}
\end{equation}

The coefficient $\chi_{\alpha k}$ measures collisions between species $\alpha$ and $k$, and $N$ denotes the number of material species, here two. The velocity differences drive momentum exchange; the temperature and relative-velocity terms describe redistribution of thermal and kinetic energy. The bulk velocity and temperature before this exchange are obtained from
\begin{equation}\label{multispecies property2}
	\begin{aligned}
		&\vec{U}_{\alpha} = \frac{1}{\rho_{\alpha}} \int \vec{u}f_\alpha {\rm d} \vec{u},\\
		&T_{\alpha}=\frac{m_\alpha}{3\rho_{\alpha}k_{B}}\int \left(\vec{u}-\vec{U}_{\alpha}\right)^{2}f_\alpha {\rm d} \vec{u}, \quad \alpha \in \mathcal{I}, \mathcal{E}.
	\end{aligned}
\end{equation}
The relaxation time $\tau_\alpha$ is calculated by
\begin{equation}\label{eq:aap-tau}
	\frac{1}{\tau_{\alpha}}=\sum_{k=\mathcal{E}}^{N}\chi_{\alpha k}n_{k},
\end{equation}
with the number density $n_\alpha$ of the species. The multispecies interaction coefficients are given by hard sphere model
\begin{equation}\label{eq:aap-chi}
	\chi_{\alpha k}=
	\frac{4\sqrt{\pi}}{3}\Bigl(\frac{2k_{B}T_{\alpha}}{m_{\alpha}}+\frac{2k_{B}T_{k}}{m_{k}}\Bigr)^{1/2}\Bigl(\frac{d_{\alpha}+d_{k}}{2}\Bigr)^{2},
\end{equation}
Here $n_\alpha=\rho_\alpha/m_\alpha$ is number density and $d_\alpha$ is the effective collision diameter. The hard-sphere interaction coefficients are used in all material tests. The relaxation time varies with density, temperature, mass, and collision diameter, thereby allowing different material transport conditions across the domain.

\subsubsection[Part 3: Radiation exchange]{Part 3: $\mathbf{W}^{**}\to\mathbf{W}^{n+1}$}
In this part, we consider that the radiation field interacts directly only with electrons. As derived in section \ref{sec:rad-ugks} by Eq.~\eqref{eq:source}, the macroscopic update is
\begin{equation}\label{eq:n+1}
\begin{aligned}
	&\partial_{t}\rho_{\mathcal E}=0,\\
	&\partial_{t}(\rho_{\mathcal E}\vec{U}_{\mathcal E})= -\vec{S}_{\mathcal{R}}(\vec{P}),\\
	&\partial_{t}E_{\mathcal E} = -cS_{\mathcal{R}}(E),
\end{aligned}
\end{equation}
During this local stage the electron density remains fixed, while radiative momentum deposition changes its bulk velocity and energy exchange changes its thermal state. The ion response is mediated by the interspecies stage.

One time step from $t^n$ to $t^{n+1}$ consists of the following stages:
\begin{enumerate}
\item Compute the material fluxes using Eq.~\eqref{eq:macro-flux-aap} and the radiation fluxes using Eqs.~\eqref{eq:micro-flux-rad} and \eqref{eq:macro-flux-rad}.
\item Update the first intermediate macroscopic state using Eq.~\eqref{eq:update-mid-1}.
\item Apply interspecies momentum and energy exchange using Eq.~\eqref{eq:aap-exchange}.
\item Apply the radiation--electron exchange in Eq.~\eqref{eq:source}.
\item Update the macroscopic state using Eq.~\eqref{eq:n+1} (or Eq.~\eqref{eq:step} for each material species) and the radiation intensity using Eq.~\eqref{eq:I}.
\end{enumerate}

\section{Numerical results}\label{sec:case}

The numerical study addresses three complementary questions: how the material solver represents species-dependent kinetic transport, how the radiation solver behaves in the diffusion regime, and how transport and exchange shape the coupled electron, ion, and radiation fields. We first use binary-mixture shocks and a Marshak wave to examine the two components. We then consider a three-temperature shock tube, radiative shocks, and the heterogeneous Tophat configuration to connect the numerical formulation to coupled physical behavior.

\subsection{Species-resolved material transport: binary-mixture shock}
We first isolate the material solver by setting aside radiation coupling and computing a binary-mixture shock at ${\rm Ma}=1.5$. This benchmark asks whether separate kinetic distributions reproduce the species-dependent density and temperature profiles obtained from a Boltzmann reference. The chosen mass ratio is an idealized mixture parameter used to test the numerical method; it does not represent the physical electron-to-ion mass ratio.

The Rankine--Hugoniot relations connect the uniform upstream state (subscript $1$) to the downstream state (subscript $2$). Here $\gamma$ is the ratio of specific heats and $n$ denotes number density. Given the upstream Mach number ${\rm Ma}_1$, the downstream state is
\begin{equation}
\begin{aligned}
	&{\rm Ma}_2= \sqrt{	\frac{{\rm Ma}_1^2(\gamma-1)+2} {2\gamma {\rm Ma}_1^2-(\gamma-1)}},\\
	&\frac{n_2}{n_1} =	\frac{(\gamma+1){\rm Ma}_1^2}	{(\gamma-1){\rm Ma}_1^2+2},\\
	&\frac{U_2}{U_1}	=	\frac{{\rm Ma}_1^2(\gamma-1)+2}	{{\rm Ma}_1^2(\gamma+1)},\\
	&\frac{T_2}{T_1} 	=	\frac{\left[2+(\gamma-1){\rm Ma}_1^2\right]		\left[2\gamma {\rm Ma}_1^2-\gamma+1\right]}	{{\rm Ma}_1^2(\gamma+1)^2}.
\end{aligned}
\end{equation}

In rarefied gas dynamics, the Knudsen number is introduced to characterize 
the degree of rarefaction. The reference Knudsen number is defined as
\begin{equation}\label{eq:kn}
	{\rm Kn}_\infty = \frac{l_\infty}{L_{\mathrm{ref}}},
\end{equation}
where the reference mean free path is 
\begin{equation*}
	l_{\infty} = \frac{1}{\sqrt{2}\pi d^2 n}.
\end{equation*}

Here $n$ is the total number density and $d$ the common effective collision diameter used to define the reference mean free path, with $d_{\mathcal{E}}=d_{\mathcal{I}}=d$. In this computation, the reference length is the cell width $L_{\mathrm{ref}}=\Delta x$, and ${\rm Kn}=2.0$. The gas is modeled with the electron--ion interaction model described in  \cite{xiao2019unified}. The velocity space for each species $\alpha$ is chosen as $u_\alpha \in (-7,7),$ according to the most probable speed of that species, and is discretized uniformly by 200 points using the midpoint rule. The mass ratio is specified as $\frac{m_{\mathcal{E}}}{m_{\mathcal{I}}} = 0.25,$
and two number density ratios are considered $ \frac{n_{\mathcal{E}}}{n_{\mathcal{I}}} = 0.1,$ and $ \frac{n_{\mathcal{E}}}{n_{\mathcal{I}}} = 0.9.$ 

\begin{figure}[H]
	\centering
	\subfloat[]{\includegraphics[width=0.45\textwidth]{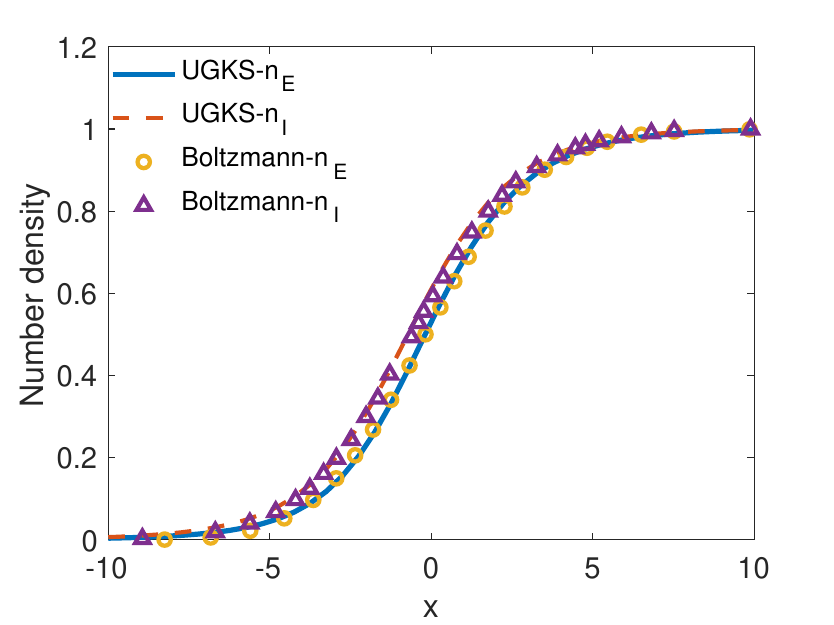}}
	\subfloat[]{\includegraphics[width=0.45\textwidth]{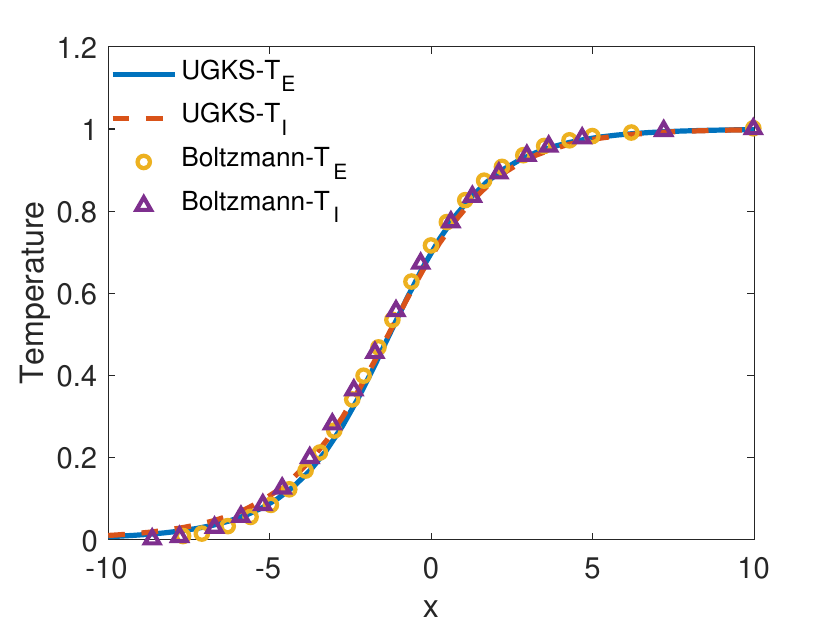}}
	\caption{Shock structure at ${\rm Ma}_\infty = 1.5$. Distributions of (a) number densities and (b) temperatures of electron and ion with the mass ratio $m_\mathcal{E}/m_\mathcal{I} = 0.25$ and number density ratio $n_\mathcal{E}/n_\mathcal{I} = 0.1$. The reference results are from \cite{kosuge2001}.}
	\label{fig:shock-chi0-1}
\end{figure}

\begin{figure}[H]
	\centering
	\subfloat[]{\includegraphics[width=0.45\textwidth]{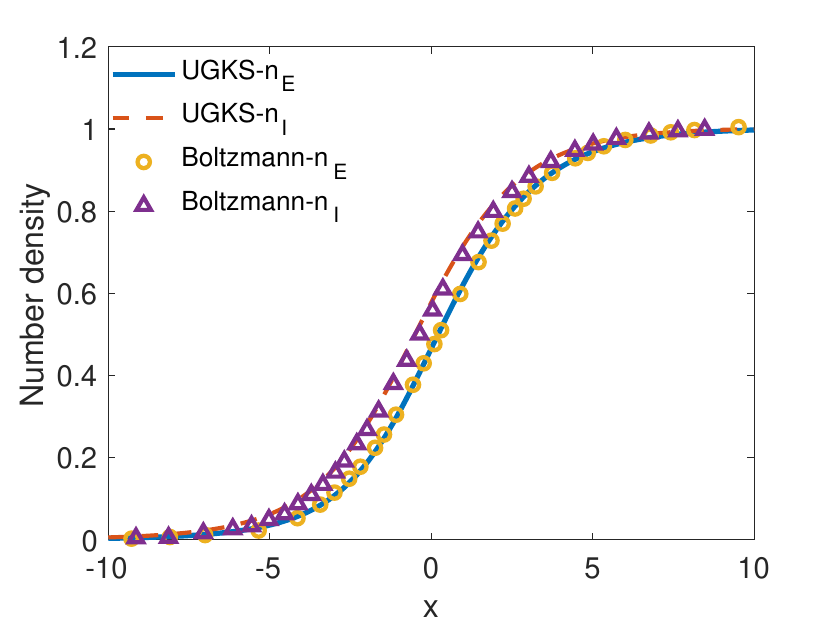}}
	\subfloat[]{\includegraphics[width=0.45\textwidth]{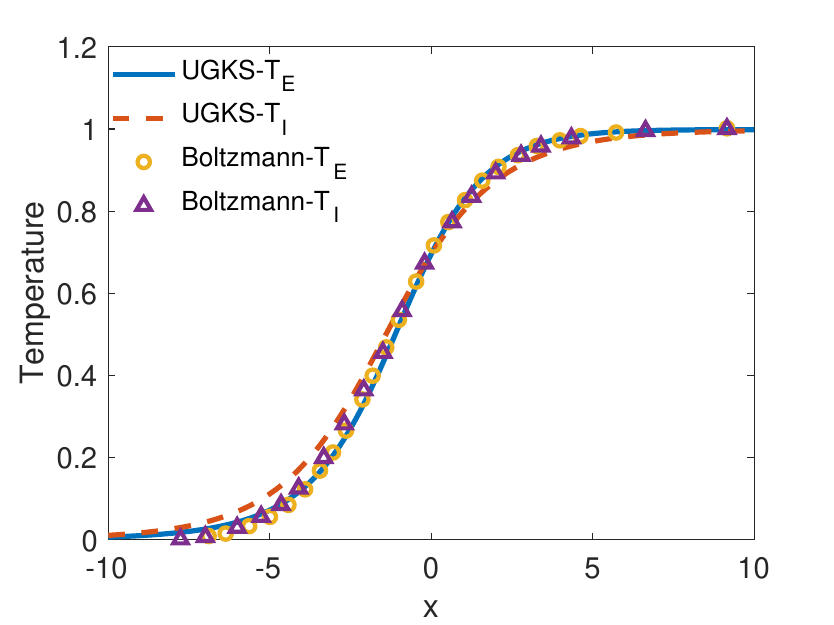}}
	\caption{Shock structure at ${\rm Ma}_\infty = 1.5$. Distributions of (a) number densities and (b) temperatures of electron and ion with the mass ratio $m_\mathcal{E}/m_\mathcal{I} = 0.25$ and number density ratio $n_\mathcal{E}/n_\mathcal{I} = 0.9$. The reference results are from \cite{kosuge2001}.}
	\label{fig:shock-chi0-9}
\end{figure}

Figures~\ref{fig:shock-chi0-1} and~\ref{fig:shock-chi0-9} compare the species number-density and temperature profiles with Boltzmann solutions computed by a numerical kernel method \cite{kosuge2001}. The separate profiles resolve how unequal particle masses and mixture composition affect the internal shock structure. This is a direct comparison of species-dependent material transport at a kinetic scale, complementing the coupled radiation cases below.

\FloatBarrier
\subsection{Radiation diffusion benchmark: Marshak wave}
\begin{figure}[!htbp]
	\centering
	\includegraphics[width=0.6\textwidth]{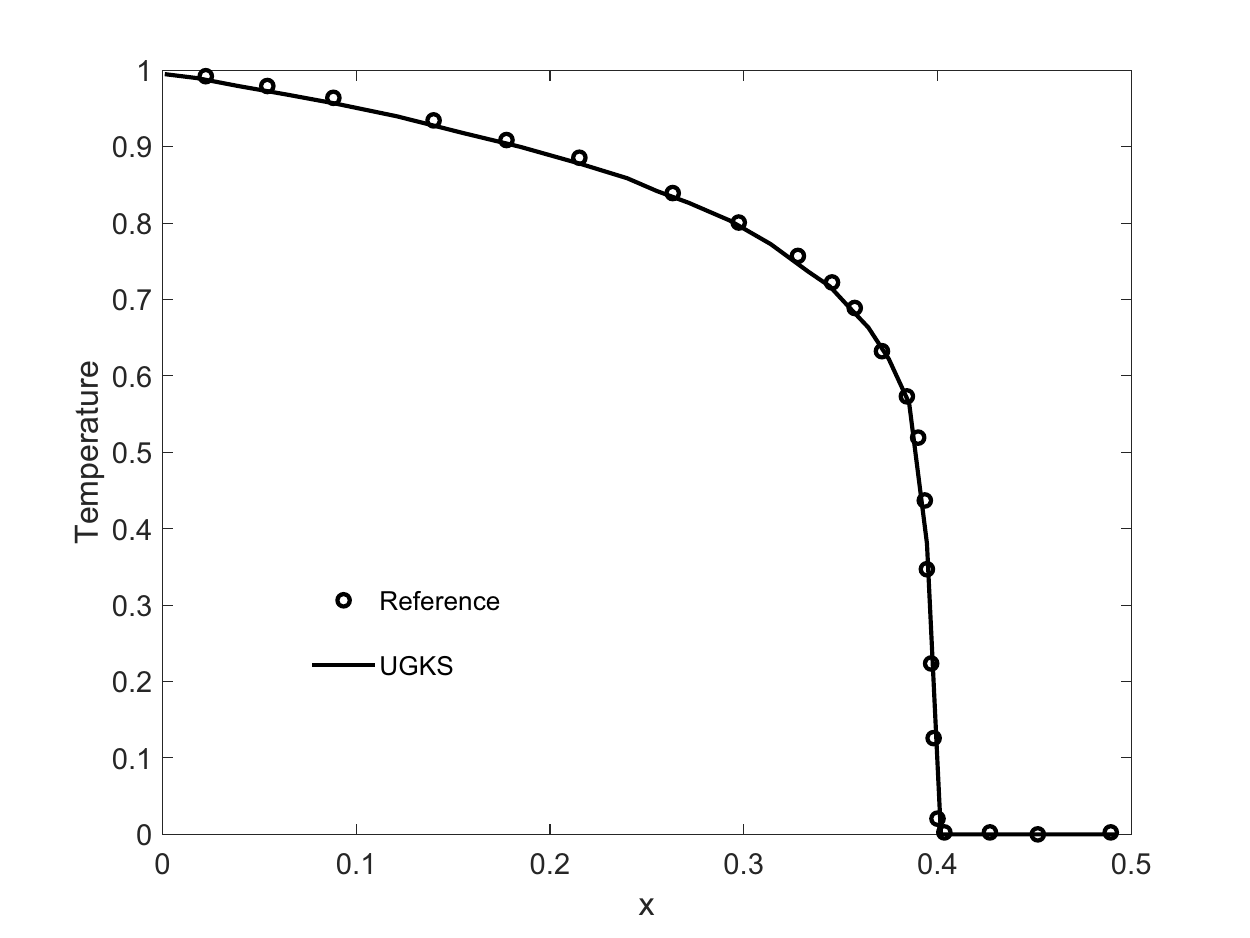}
	\caption{Marshak-wave temperature profile in the diffusion regime with $\sigma=100/T^{3}$. The UGKS result is compared with the diffusion reference of \cite{sun2015gray}.}
	\label{fig:rad-gray}
\end{figure}

We next test the radiative component against a Marshak-wave reference in an optically thick, stationary material. Radiation heats an initially cold medium while the material velocity is held at zero. This setup examines radiative diffusion and energy exchange under the specified conditions. The configuration follows the benchmark used in our earlier study \cite{xu2025}. The absorption/emission coefficient is $\sigma = 100/T^{3}$. The initial electron temperature is $T_{\mathcal{E}}=10^{-4}$. The radiation intensity is initially zero for $x\in(0,1)$, with boundary values $I(0)=1$ and $I(1)=0$. We use 200 uniform spatial cells and 30 midpoint angular quadrature points over $\Omega_x\in(-1,1)$. Figure~\ref{fig:rad-gray} compares the computed temperature profile with the diffusion reference \cite{sun2015gray}. The comparison checks whether replacing the material transport solver preserves the expected thick-medium behavior in this benchmark.

\FloatBarrier
\subsection{Three-temperature double-Lax shock tube}

\begin{figure}[!htbp]
	\centering
	\subfloat[]{\includegraphics[width=0.33\textwidth]{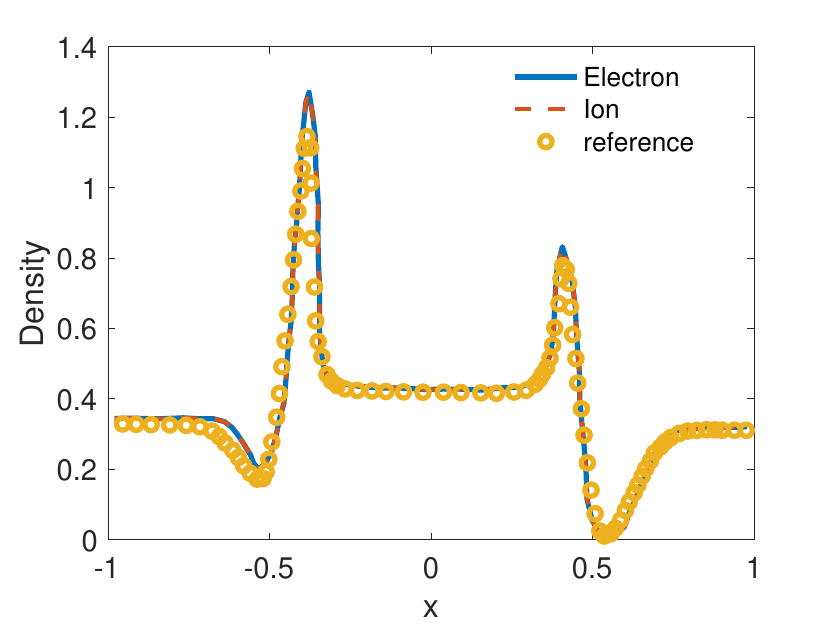}}
	\subfloat[]{\includegraphics[width=0.33\textwidth]{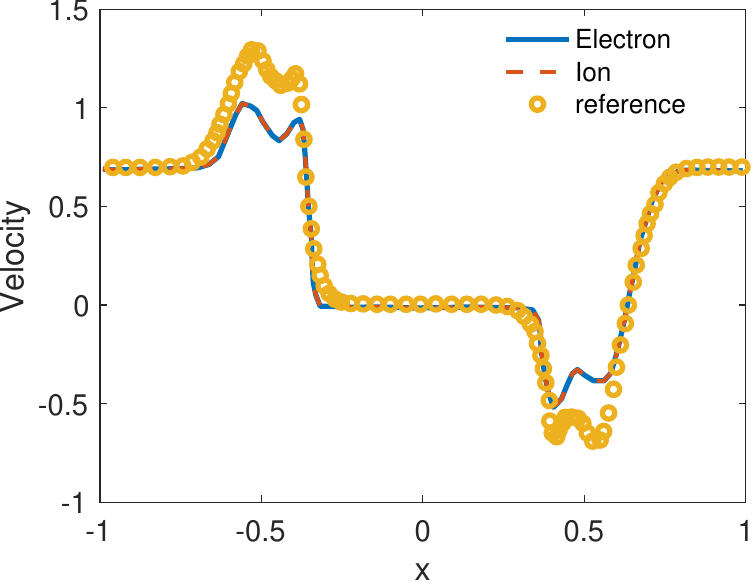}}
	\subfloat[]{\includegraphics[width=0.33\textwidth]{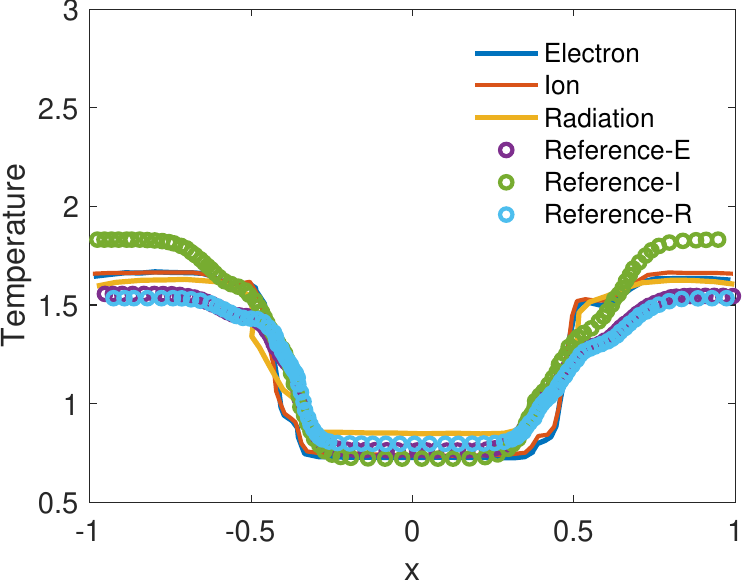}} 
	\caption{Three-temperature double-Lax shock tube at ${\rm Kn}=10^{-5}$, $m_\mathcal{I}/m_\mathcal{E}=1$, and $n_\mathcal{I}/n_\mathcal{E}=1$. Panels show (a) densities, (b) velocities, and (c) electron, ion, and radiation temperatures, compared with the radiation-hydrodynamic three-temperature reference of \cite{cheng2024}.}
	\label{fig:double lax}
\end{figure}

This problem compares the coupled species and radiation profiles with the three-temperature continuum calculation of \cite{cheng2024}. It connects the kinetic formulation to a continuum benchmark at ${\rm Kn}=10^{-5}$ and examines the resulting density, velocity, and thermal wave structures. The nondimensional initial condition is
\begin{equation*}
	\left\{\begin{array}{l l l l}
		{\rho=0.445,\quad u=0.698,\quad p=1.176, \quad -1\leq x\leq-0.5}\\
		{\rho=0.5,  \qquad u=0,   ~~ \qquad p =0.19, \quad -0.5\leq x\leq0.5}\\
		{\rho=0.445,\quad u=0.698,\quad p=1.176, \quad 0.5\leq x\leq1}
	\end{array}\right.
\end{equation*}
with
\begin{equation*}
	\rho = \rho_{\mathcal{E}} = \rho_{\mathcal{I}}, \quad
	u = U_{\mathcal{E}}, \quad
	p = p_{\mathcal{E}} = p_{\mathcal{I}},
\end{equation*}
Electron and ion collisions are modeled with the hard-sphere model. The molecular diameter $d = d_\mathcal{E} = d_\mathcal{I} $ and number density $n = n_\mathcal{E} + n_\mathcal{I} $ are set for the definition of Knudsen number. The reference length defining the Knudsen number is $L_{ref}=1$, and the spatial domain is discretized into 200 uniform cells. The velocity space is discretized by 100 mesh points using the midpoint rule with the range $u_\alpha \in (-7,7)$ according to the most probable speed of each species $\alpha$. Results are reported at $t=1$.
The reference formulation evolves a common material velocity and three internal energies with diffusion and thermal relaxation \cite{cheng2024}. The present scheme instead transports separate electron and ion distributions and the angular radiation intensity, with species momentum exchange and velocity-dependent radiation--electron sources. Figure~\ref{fig:double lax} shows the corresponding wave patterns: density features occur in similar locations, while differences are visible in the velocity extrema and temperature profiles. The separate electron and ion velocities and temperatures expose the species dynamics retained by the coupled formulation. This comparison links those additional degrees of freedom to the continuum wave pattern and identifies the velocity and thermal fields as sensitive observables.

\FloatBarrier
\subsection[Radiative shocks at Mach 1.5 and 3.0]{Radiative shocks at $\mathrm{Ma}=1.5$ and $\mathrm{Ma}=3.0$}
Radiative shocks connect material compression to radiative heating and thermal relaxation. We consider ${\rm Ma}=1.5$ and ${\rm Ma}=3.0$ to examine the change from a moderate shock to a stronger shock with a temperature overshoot. The configurations follow \cite{bolding2017second,lowrie2008radiative,xu2025}. The radiation intensity is initialized according to the electron temperature. The spatial domain is partitioned into 200 uniformly distributed cells and 30 midpoint angular quadrature points over $\Omega_x\in(-1,1)$. We consider $\gamma=5/3$ and adopt the grey approximation. All variables are nondimensionalized by the upstream state $(\rho_0,T_0)$, and radiation quantities are scaled by $a_R T_0^4$. The upstream state is initialized as $\rho_L=1$, $T_{e,L}=T_{i,L}=1$, $u_L=\mathrm{Ma}$, with isotropic radiation in local equilibrium with the electron and ion temperature, i.e. $E_{r,L}=T_{e,L}^4$ and $F_{r,L}=0$. The profiles exhibit how compression and radiation exchange together determine the material and radiation temperatures.

\begin{figure}[!htbp]
	\centering
	\subfloat[]{\includegraphics[width=0.33\textwidth]{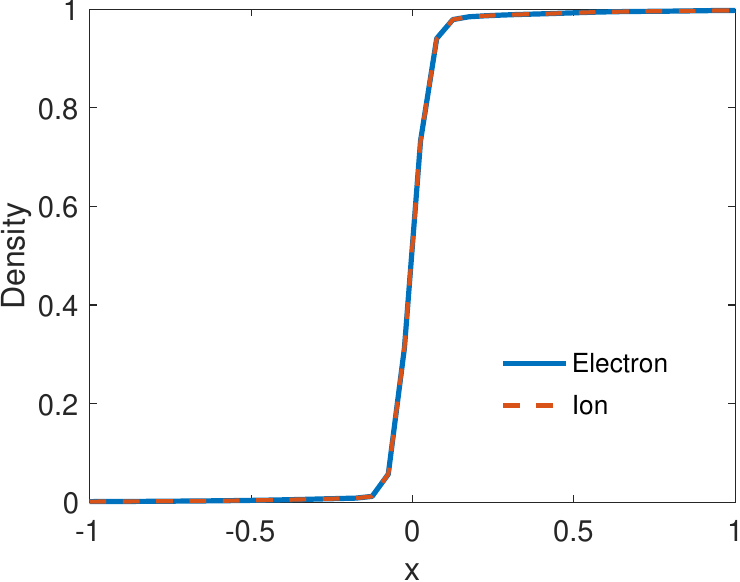}}
	\subfloat[]{\includegraphics[width=0.33\textwidth]{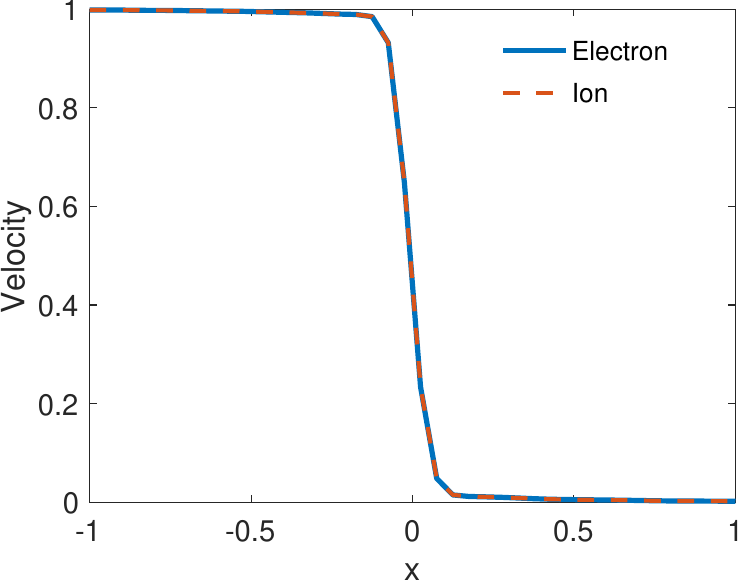}}
	\subfloat[]{\includegraphics[width=0.33\textwidth]{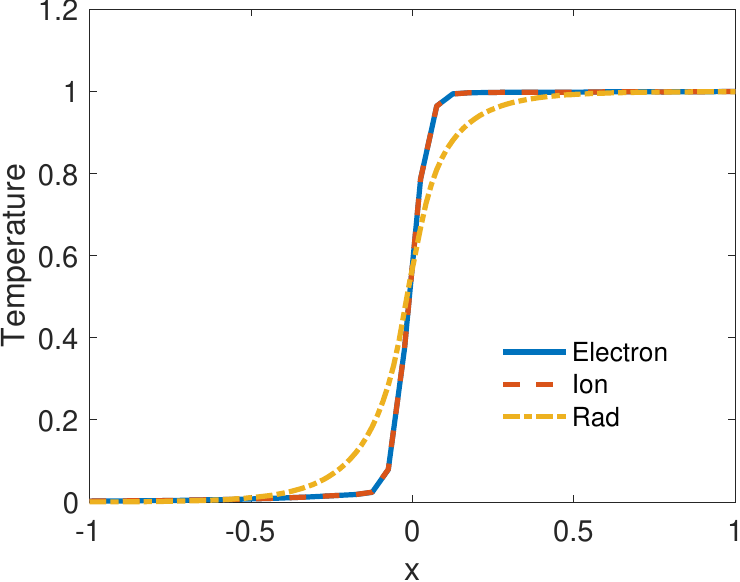}}
	\caption{Radiative shock at $\mathrm{Ma}=1.5$ and $m_\mathcal{E}/m_\mathcal{I}=1$: (a) normalized density, (b) normalized velocity, and (c) normalized material and radiation temperatures.}
	\label{fig:radshock-1.5-2}
\end{figure}

\begin{figure}[!htbp]
	\centering
	\subfloat[]{\includegraphics[width=0.33\textwidth]{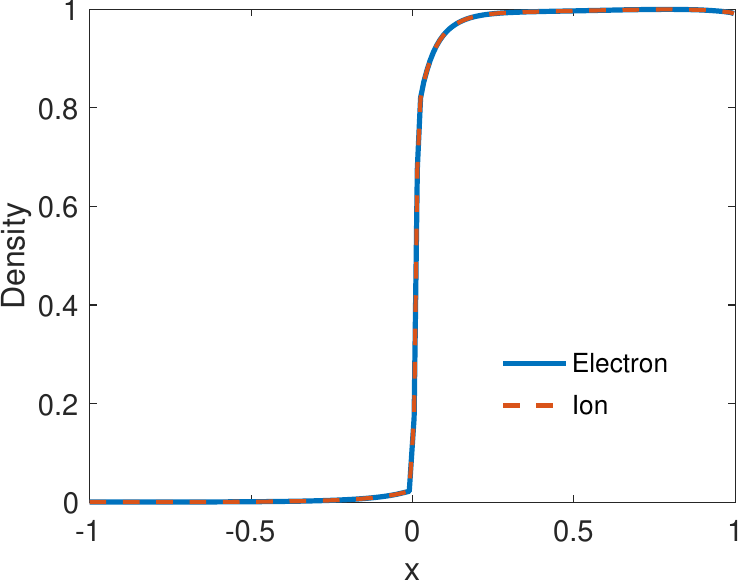}}
	\subfloat[]{\includegraphics[width=0.33\textwidth]{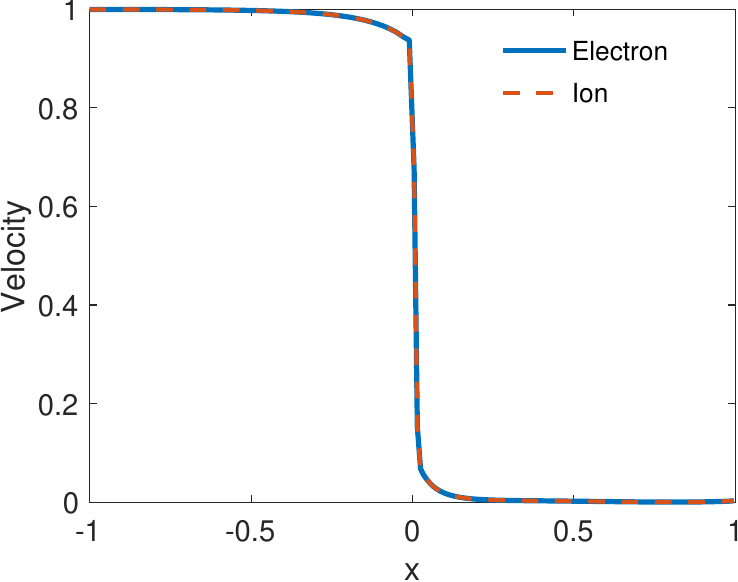}}
	\subfloat[]{\includegraphics[width=0.33\textwidth]{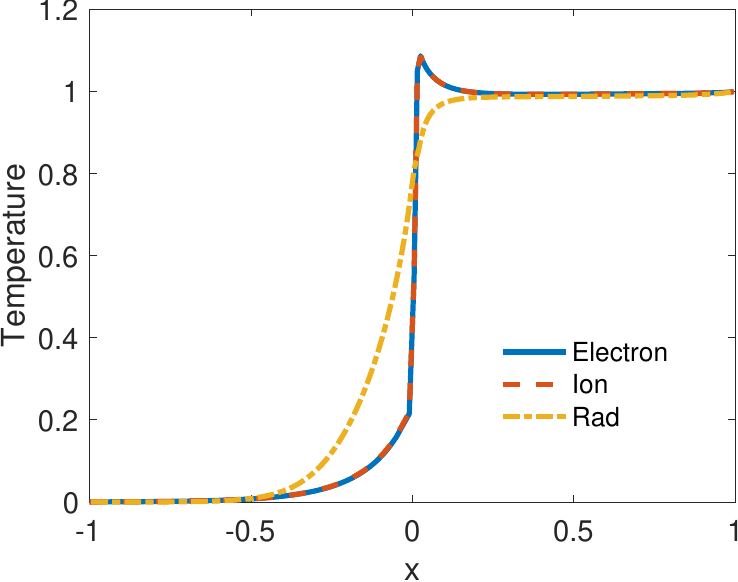}}
	\caption{Radiative shock at $\mathrm{Ma}=3.0$ and $m_\mathcal{E}/m_\mathcal{I}=1$: (a) normalized density, (b) normalized velocity, and (c) normalized material and radiation temperatures.}
	\label{fig:radshock-3.0-2}
\end{figure}

Figures~\ref{fig:radshock-1.5-2} and~\ref{fig:radshock-3.0-2} display profiles normalized by their far-field states. Density and temperature are scaled to zero upstream and unity downstream, while velocity is scaled to unity upstream and zero downstream. These plotting transformations are applied after the physical nondimensionalization described above.

 Figure~\ref{fig:radshock-1.5-2} shows electron and ion density, velocity, and temperature, together with radiation temperature, for $\mathrm{Ma}=1.5$ and $m_\mathcal{E}/m_\mathcal{I}=1$. The material temperature exhibits a jump at the hydrodynamic shock, and its maximum does not exceed the far-downstream temperature. 
These results are consistent with the solutions in \cite{lowrie2008radiative,bolding2017second}. Figure~\ref{fig:radshock-3.0-2} shows the corresponding profiles for $\mathrm{Ma}=3.0$ and the same mass ratio. In this stronger shock, the material temperature peaks just downstream of the hydrodynamic jump, forming a Zel'dovich spike as reported in \cite{lowrie2008radiative,bolding2017second}. The transition from the moderate-shock profile to the stronger-shock overshoot demonstrates the coupled description of compression, radiation transport, and thermal exchange.

\FloatBarrier
\subsection{Tophat configuration}
\begin{figure}[!htbp]
	\centering
	\includegraphics[width=0.8\textwidth]{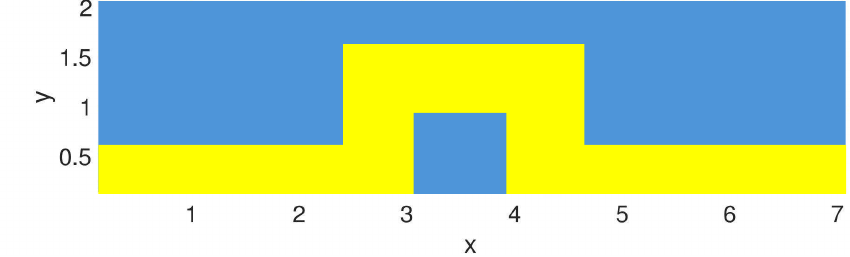}
	\caption{Initial configuration of the two-dimensional Tophat problem. The blue region corresponds to the optically thick and continuum flow regime, while the yellow region represents the optically thin and free-molecular flow regime. }
	\label{fig:demo-tophat1}
\end{figure}

The two-dimensional Tophat problem places optically thick material next to dilute, optically thin regions. Its geometry combines changes in material rarefaction and photon interaction length with spatially nonuniform heating. This provides a setting for examining the interplay of species transport, radiative transport, and thermal nonequilibrium. Extending the configuration of \cite{xu2025} to a kinetic material description allows electron and ion distributions to evolve through both dense and dilute regions. 
As shown in Fig.~\ref{fig:demo-tophat1}, the initial condition is set in the optically thick continuum-flow regime with $\sigma = 10^{2}, \rho = 1.0, p = 1.0,$ in $[x_1,x_2]\times[y_1,y_2] \in \{[3.0,4.0]\times[0,1.0], [0,2.5]\times[0.5,2.0], [4.5,7.0]\times[0.5,2.0], [2.5,4.5]\times[1.5,2.0]\}$, and an optically thin and free molecular flow regime with $\sigma = 10^{-1}, \rho = 10^{-5}, p = 10^{-5}$ otherwise. The mass ratio and number density ratio of electron and ion are $rM = \frac{m_{\mathcal{I}}}{m_\mathcal{E}}  = 4.0, \frac{n_{\mathcal{I}}}{n_\mathcal{E}+n_\mathcal{I}} = 0.2$.

The Knudsen number is defined by Eq.~\eqref{eq:kn}, where $L_{ref} = 1$, and the reference mean free path is calculated by the number density and molecular diameter for mean free path as $n = n_{\mathcal{I}} + n_{\mathcal{E}}$ and $d = d_\mathcal{I} = d_\mathcal{E}$. The physical space $x \times  y = 7 \times 2$ is discretized by $70 \times 20$ uniform meshes. The velocity space is discretized by $30\times30$ points using the midpoint rule with the range $u_{\alpha} \in [-7,7]$ according to the most probable speed of each species $\alpha$.

\begin{figure}[!htbp]
	\centering
	\subfloat[]{\includegraphics[width=0.5\textwidth]{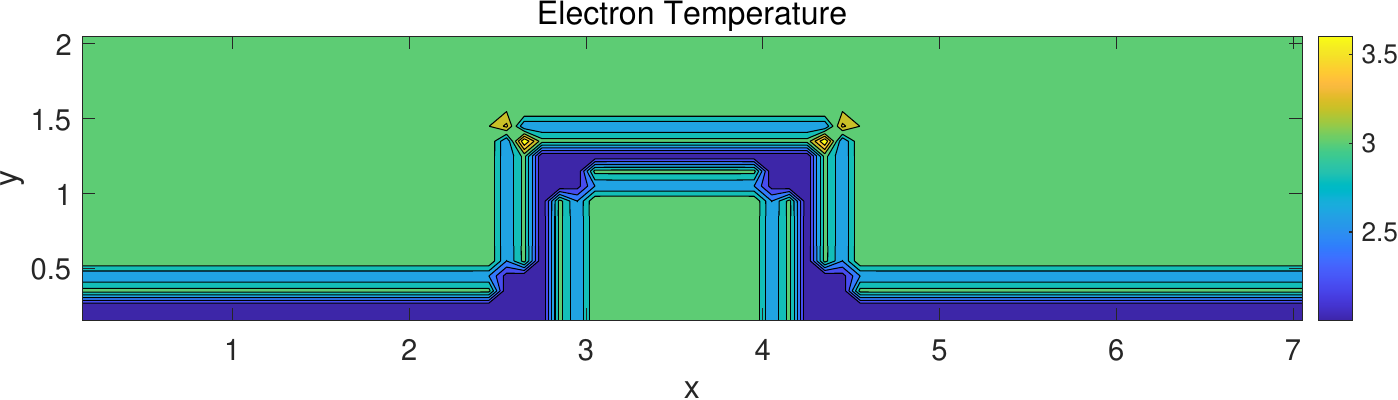}}	\subfloat[]{\includegraphics[width=0.5\textwidth]{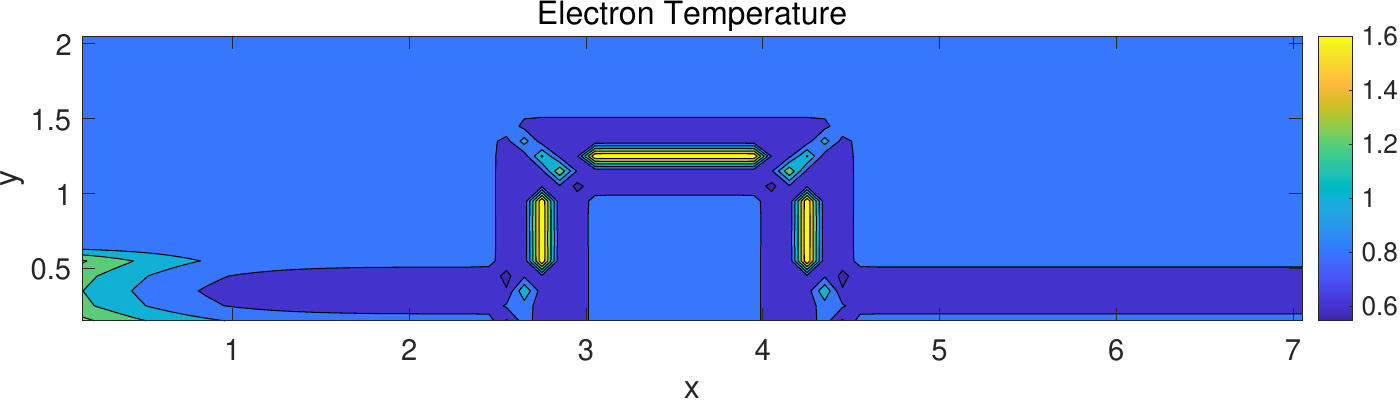}}
	
	\subfloat[]{\includegraphics[width=0.5\textwidth]{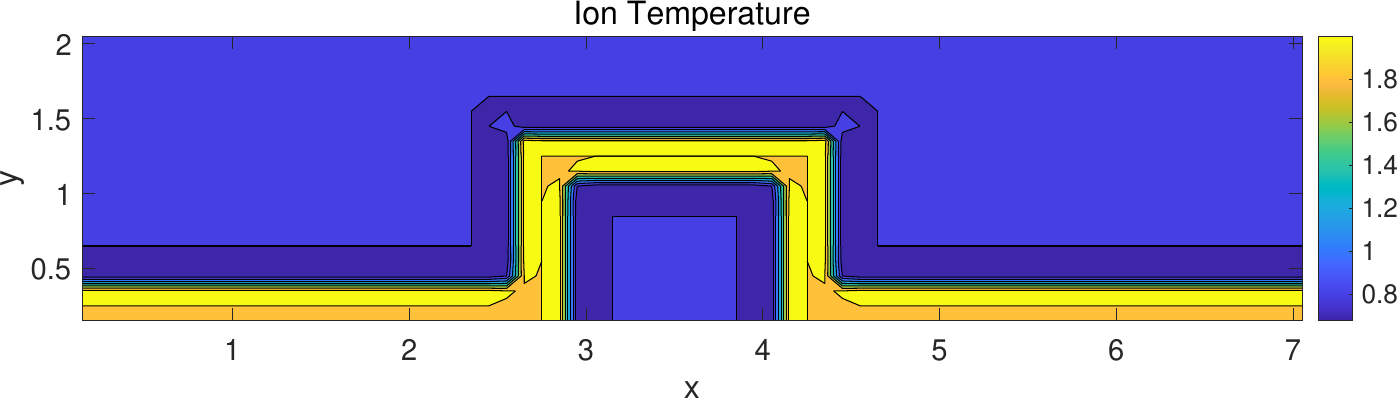}}	\subfloat[]{\includegraphics[width=0.5\textwidth]{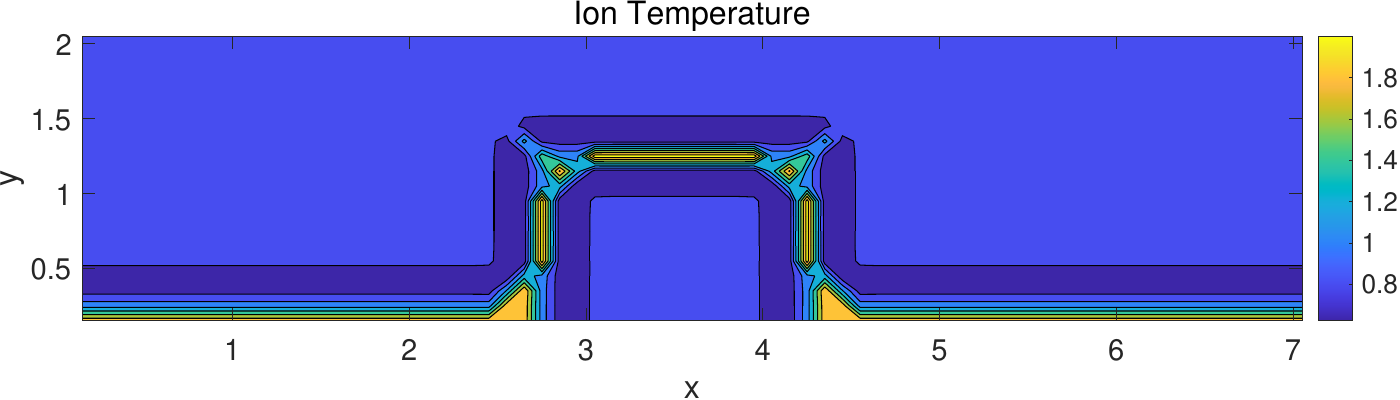}}
	
	\subfloat[]{\includegraphics[width=0.5\textwidth]{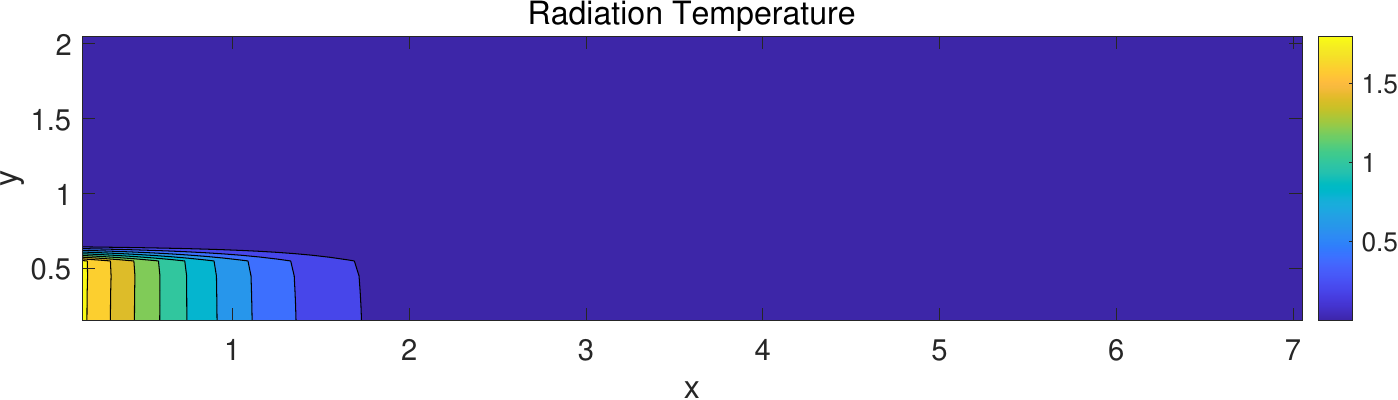}}
	\subfloat[]{\includegraphics[width=0.5\textwidth]{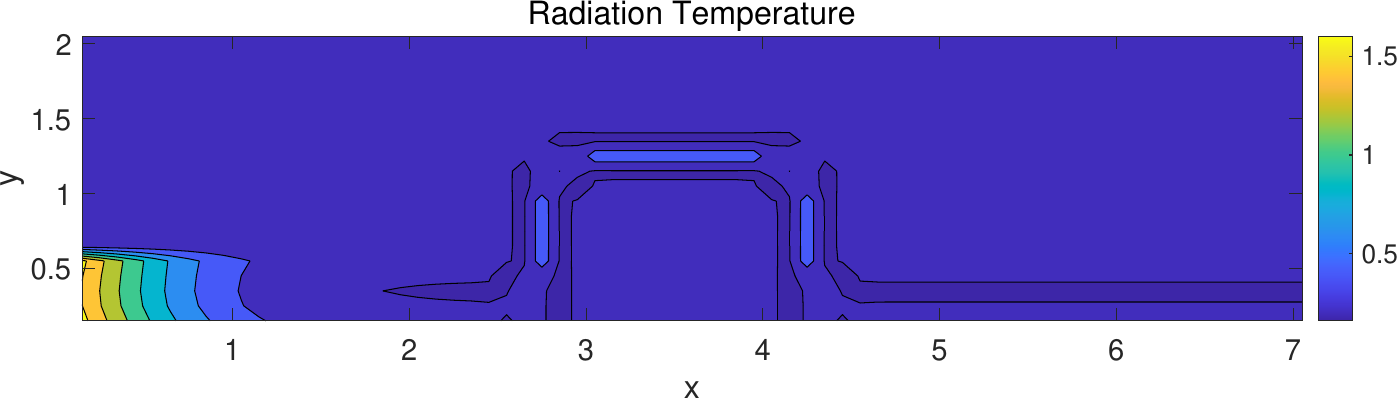}}
	\caption{Two-dimensional Tophat at $t=0.01$ with $m_\mathcal{I}/m_\mathcal{E}=4$ and $n_\mathcal{I}/(n_\mathcal{E}+n_\mathcal{I})=0.2$. Rows show electron, ion, and radiation temperatures, respectively. The left column omits radiation--material exchange; the right column includes it.}
	\label{fig:tophat-0.01-1}
\end{figure}

\begin{figure}[!htbp]
	\centering
	\subfloat[]{\includegraphics[width=0.45\textwidth]{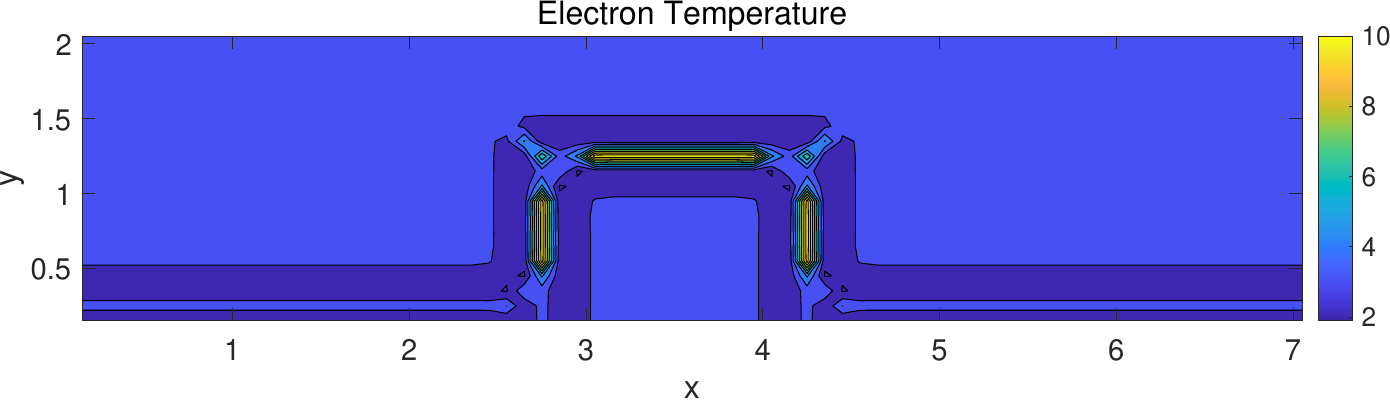}}	\subfloat[]{\includegraphics[width=0.45\textwidth]{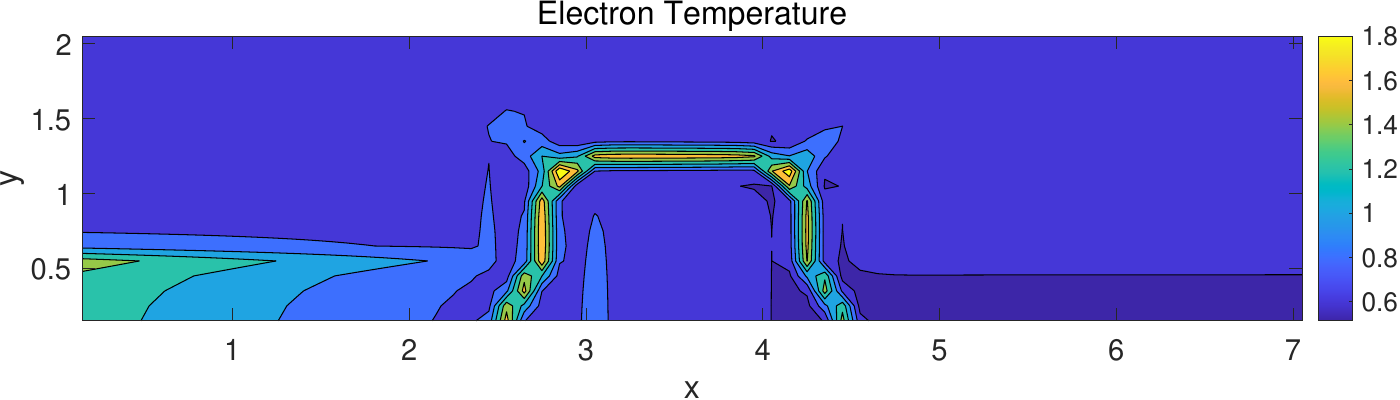}}
	
	\subfloat[]{\includegraphics[width=0.45\textwidth]{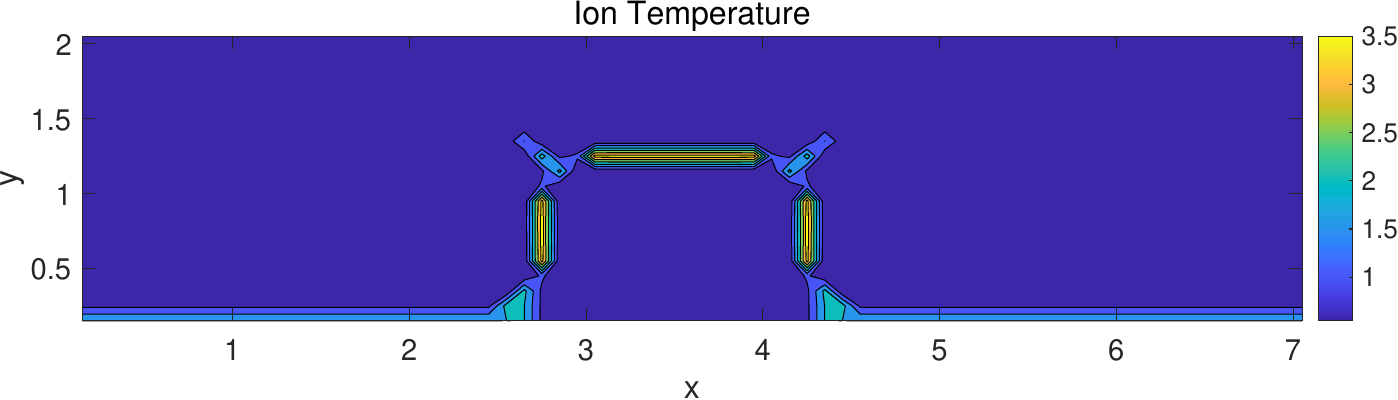}}	\subfloat[]{\includegraphics[width=0.45\textwidth]{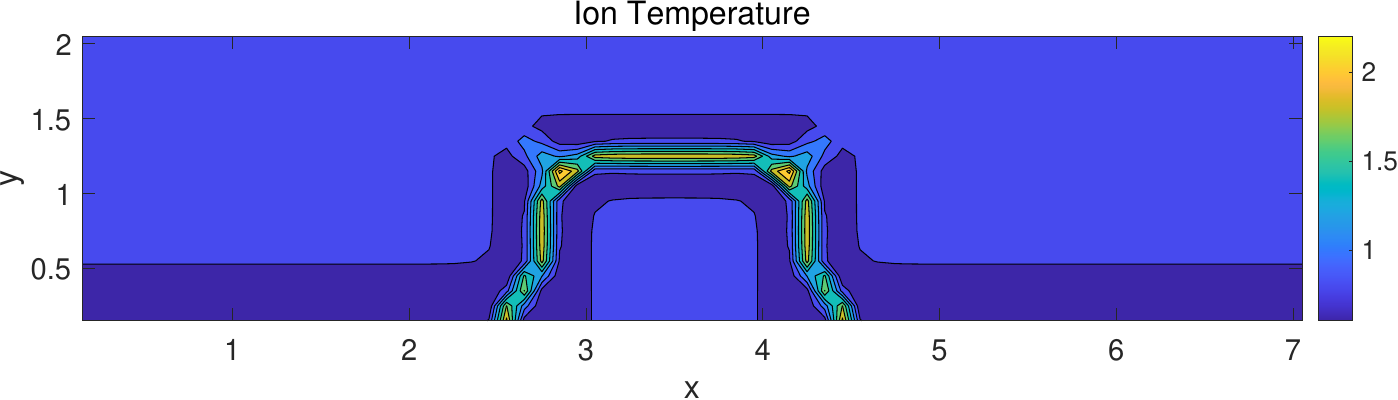}}
	
	\subfloat[]{\includegraphics[width=0.45\textwidth]{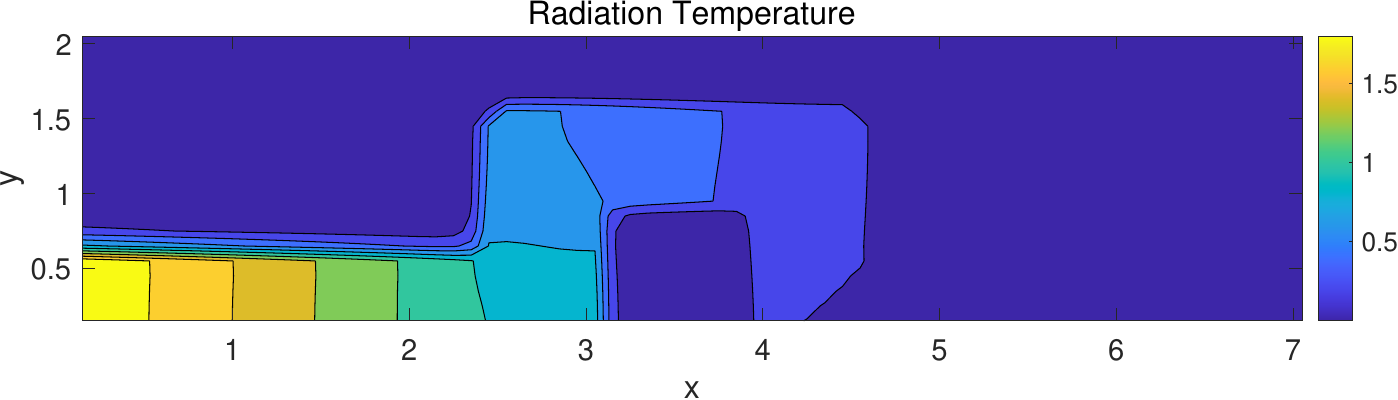}}
	\subfloat[]{\includegraphics[width=0.45\textwidth]{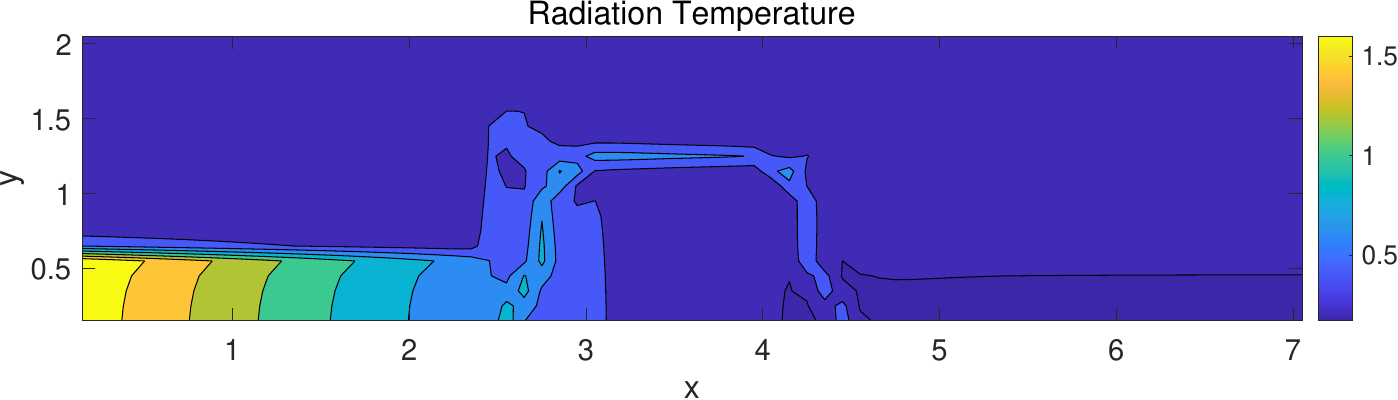}}
	\caption{Two-dimensional Tophat at $t=0.1$ with $m_\mathcal{I}/m_\mathcal{E}=4$ and $n_\mathcal{I}/(n_\mathcal{E}+n_\mathcal{I})=0.2$. Rows show electron, ion, and radiation temperatures, respectively. The left column omits radiation--material exchange; the right column includes it.}
	\label{fig:tophat-0.1-1}
\end{figure}

\begin{figure}[!htbp]
	\centering
	\subfloat[]{\includegraphics[width=0.45\textwidth]{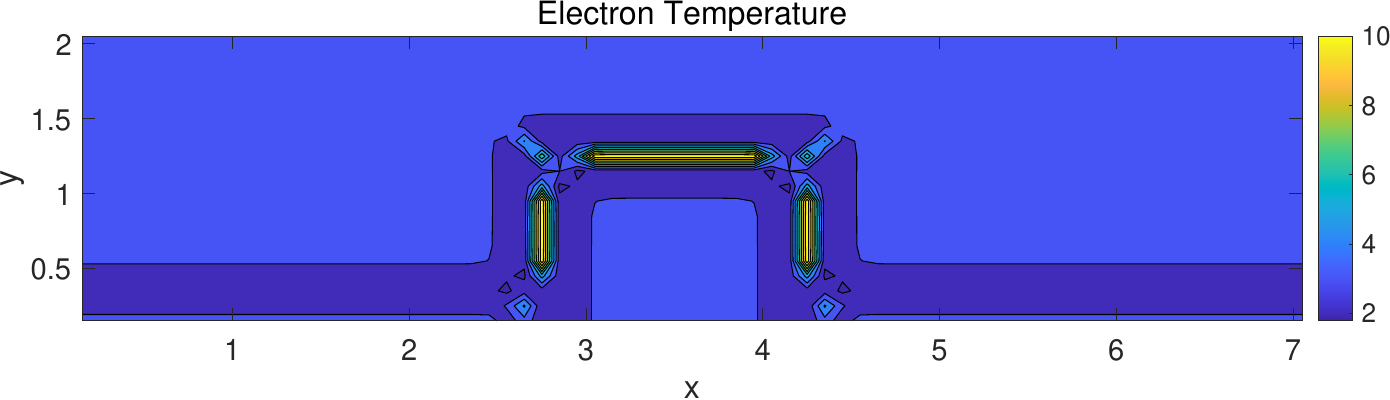}}	\subfloat[]{\includegraphics[width=0.45\textwidth]{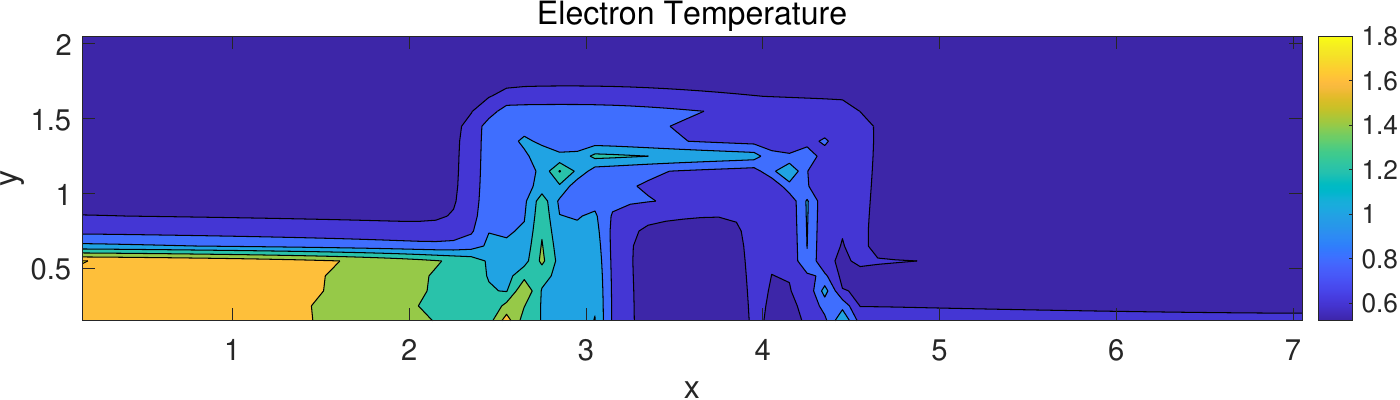}}
	
	\subfloat[]{\includegraphics[width=0.45\textwidth]{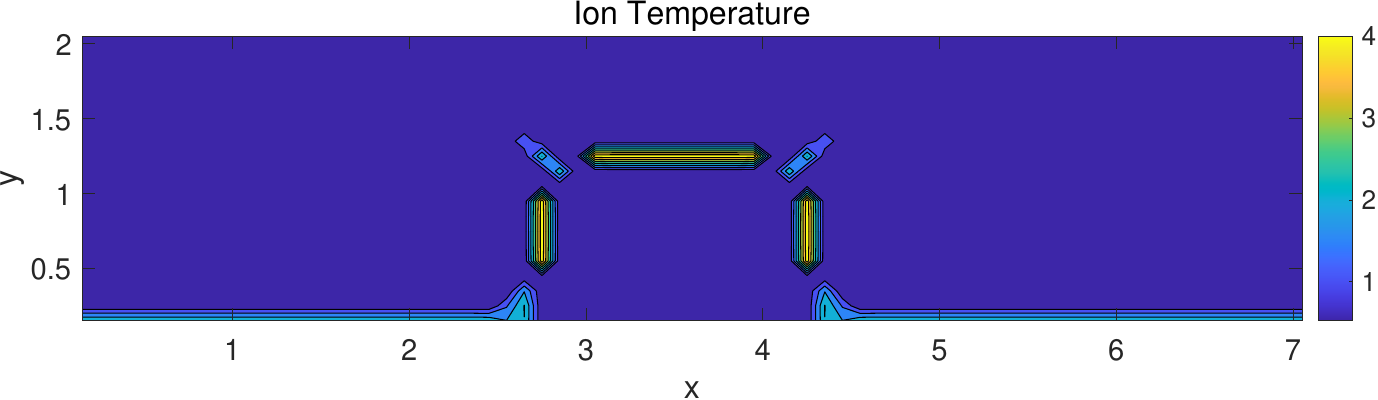}}	\subfloat[]{\includegraphics[width=0.45\textwidth]{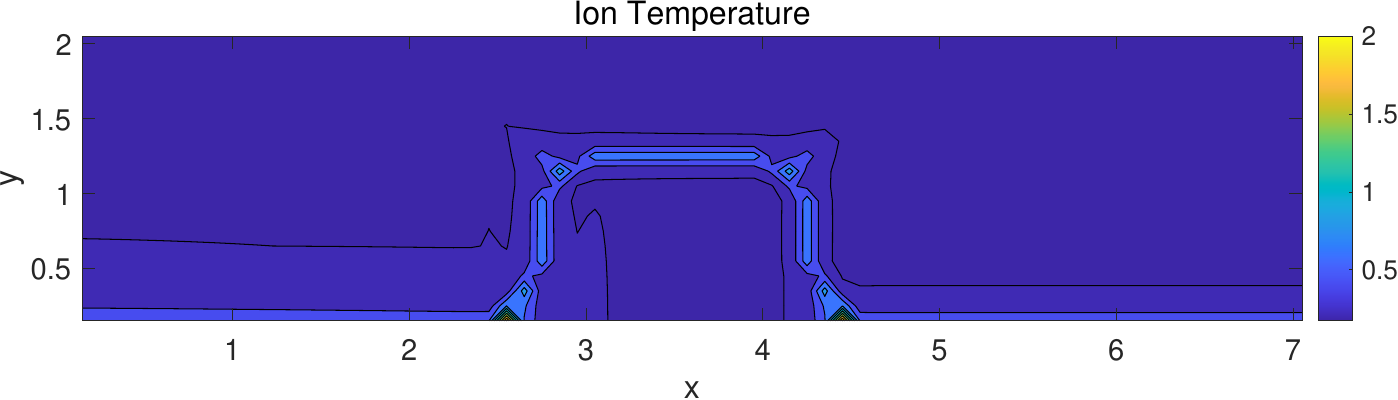}}
	
	\subfloat[]{\includegraphics[width=0.45\textwidth]{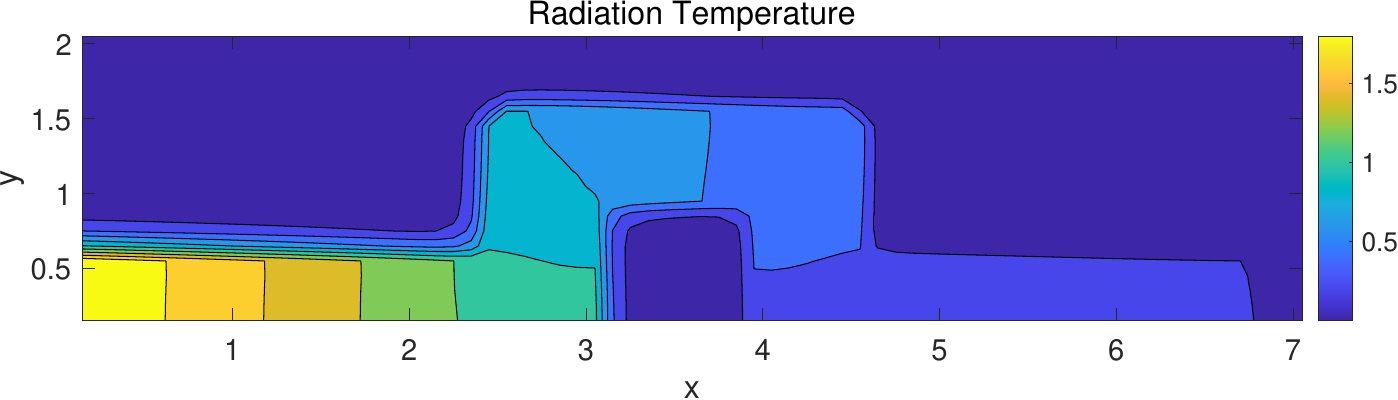}}
	\subfloat[]{\includegraphics[width=0.45\textwidth]{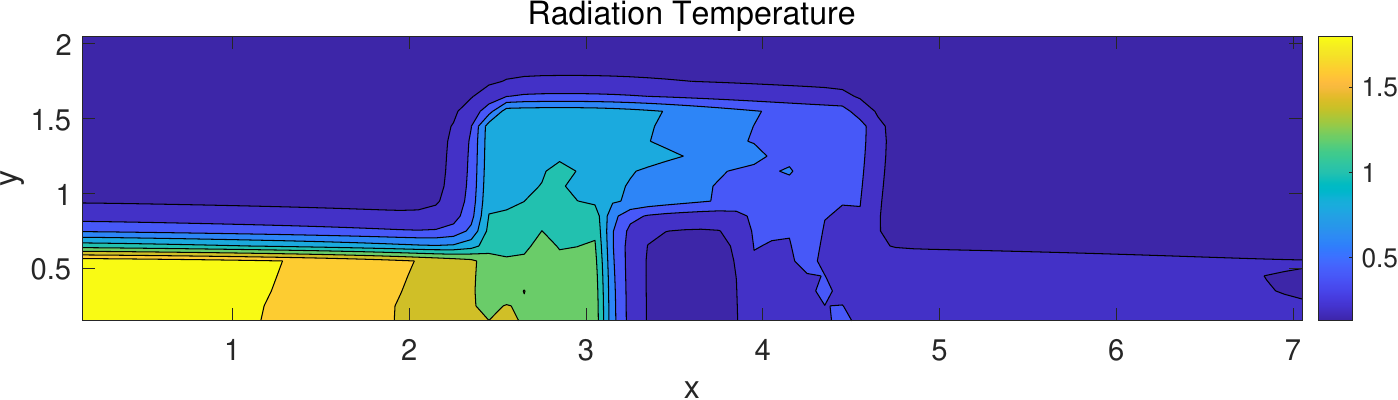}}
	\caption{Two-dimensional Tophat at $t=0.2$ with $m_\mathcal{I}/m_\mathcal{E}=4$ and $n_\mathcal{I}/(n_\mathcal{E}+n_\mathcal{I})=0.2$. Rows show electron, ion, and radiation temperatures, respectively. The left column omits radiation--material exchange; the right column includes it.}
	\label{fig:tophat-0.2-1}
\end{figure}

\begin{figure}[!htbp]
	\centering
	\subfloat[]{\includegraphics[width=0.45\textwidth]{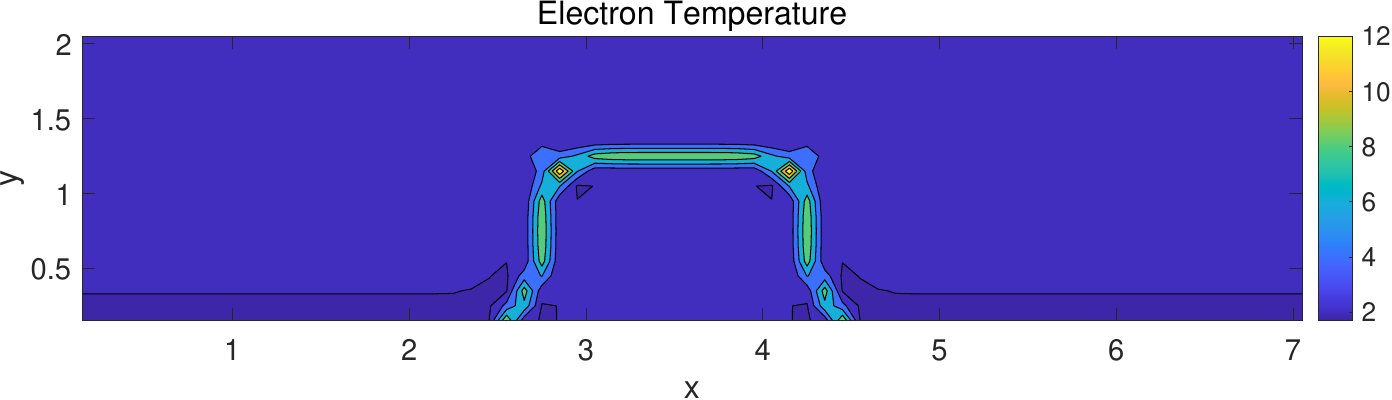}}	\subfloat[]{\includegraphics[width=0.45\textwidth]{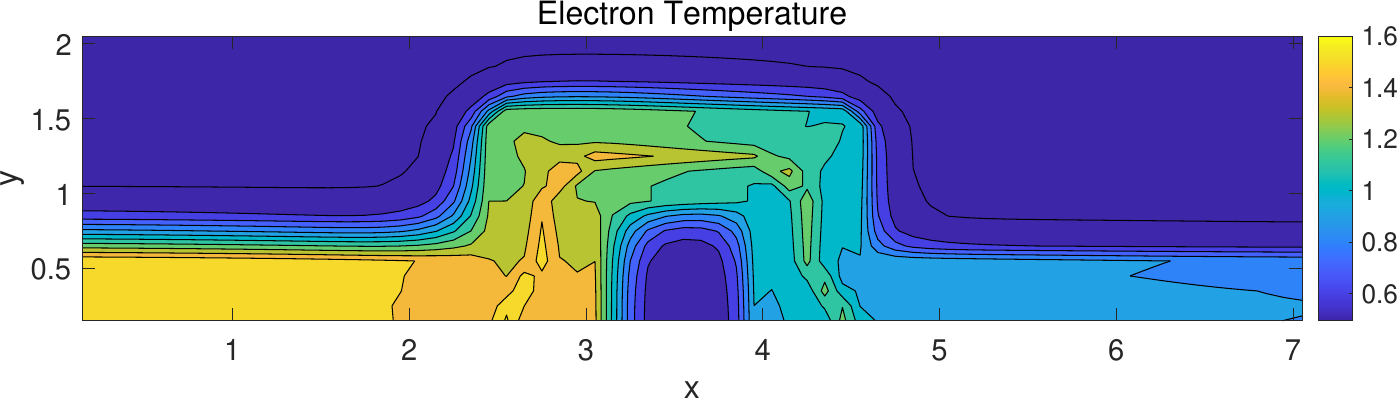}}
	
	\subfloat[]{\includegraphics[width=0.45\textwidth]{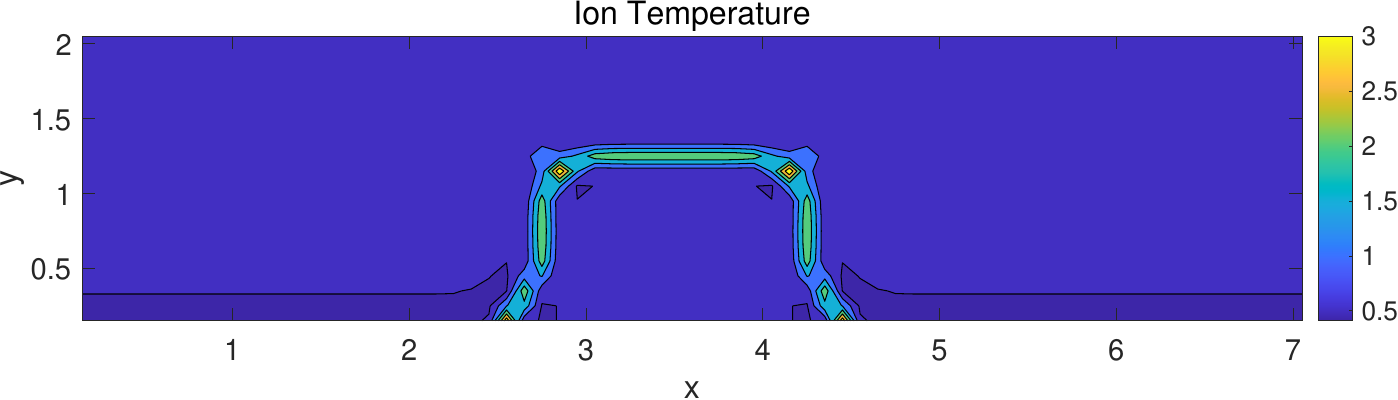}}	\subfloat[]{\includegraphics[width=0.45\textwidth]{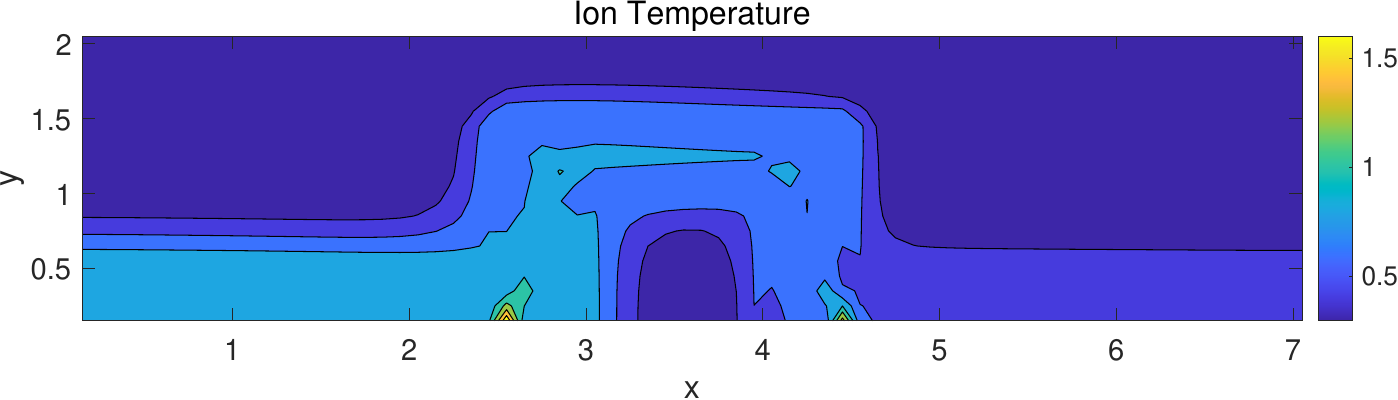}}
	
	\subfloat[]{\includegraphics[width=0.45\textwidth]{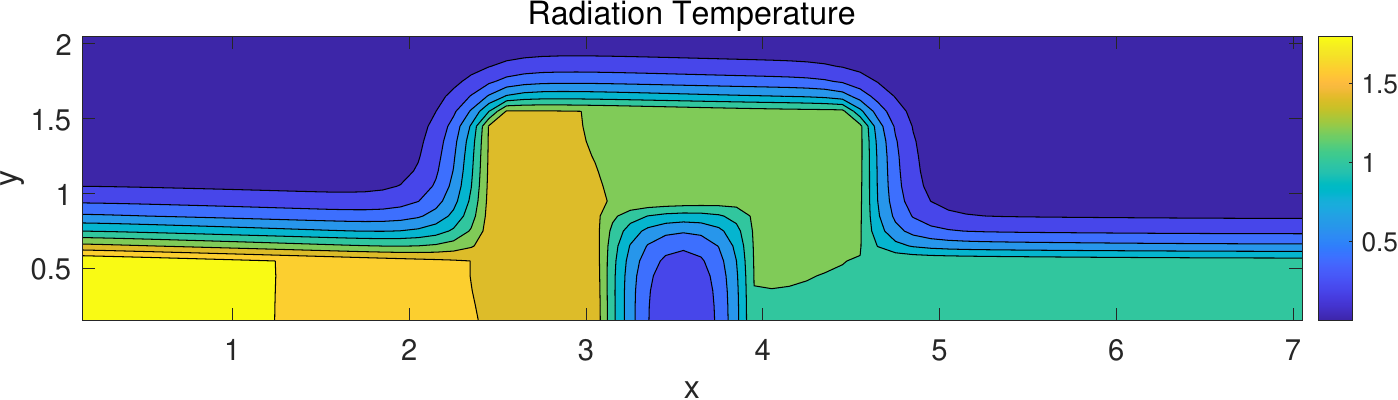}}
	\subfloat[]{\includegraphics[width=0.45\textwidth]{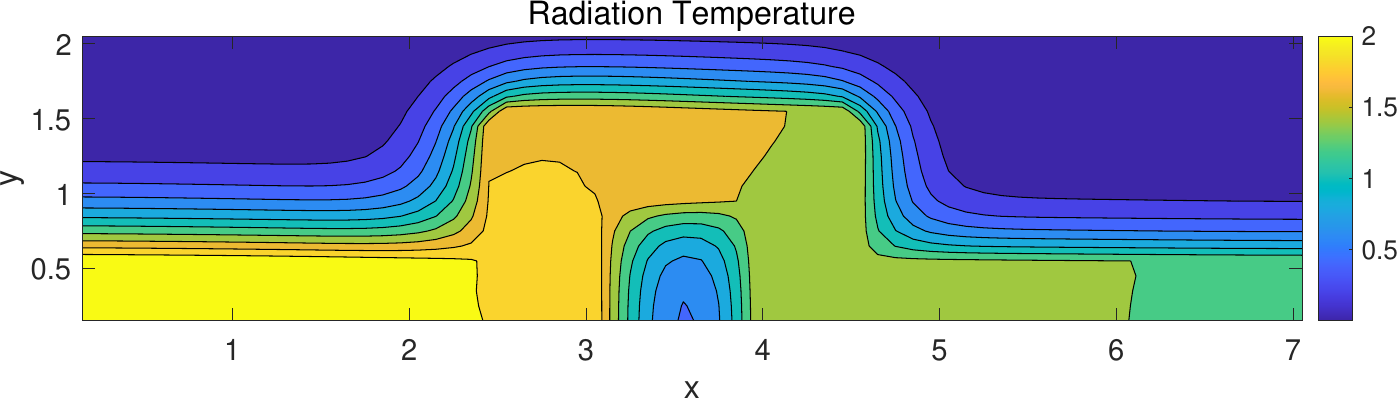}}
	\caption{Two-dimensional Tophat at $t=1.0$ with $m_\mathcal{I}/m_\mathcal{E}=4$ and $n_\mathcal{I}/(n_\mathcal{E}+n_\mathcal{I})=0.2$. Rows show electron, ion, and radiation temperatures, respectively. The left column omits radiation--material exchange; the right column includes it.}
	\label{fig:tophat-1.0-1}
\end{figure}

The snapshots at $t=0.01$, $0.1$, $0.2$, and $1.0$ compare evolution with radiation--material exchange suppressed (left column) and included (right column). The material, radiation, and geometric settings are otherwise the same. The paired runs therefore reveal the influence of the included exchange on the species temperature fields. In particular, the $t=0.2$ snapshots show localized material hot regions in the uncoupled calculation and a broader temperature response with coupling. Electron, ion, and radiation temperatures have distinct spatial patterns, making thermal nonequilibrium visible in the heterogeneous domain. Electrons respond directly to radiation, and electron--ion collisions redistribute the exchanged momentum and energy. Each panel has its own color scale; the color bars should be used when comparing temperature magnitudes.

\begin{figure}[!htbp]
	\centering
	\subfloat[]{\includegraphics[width=0.33\textwidth]{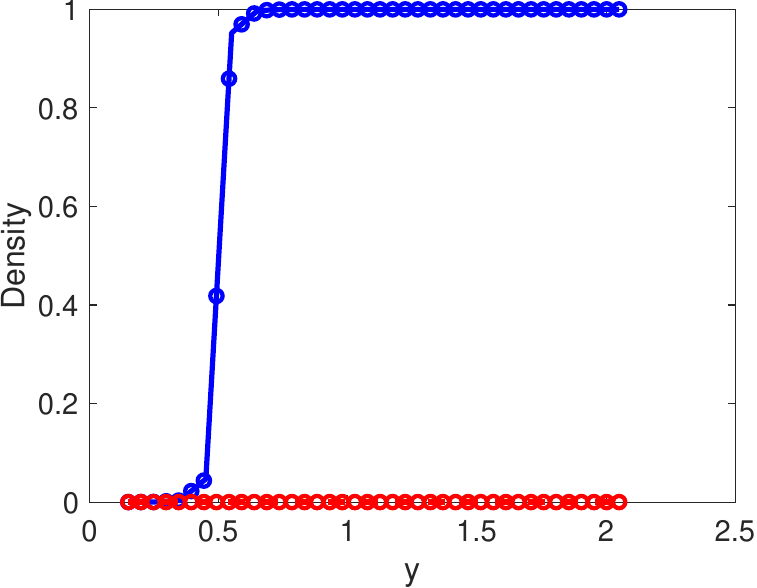}}
	\subfloat[]{\includegraphics[width=0.33\textwidth]{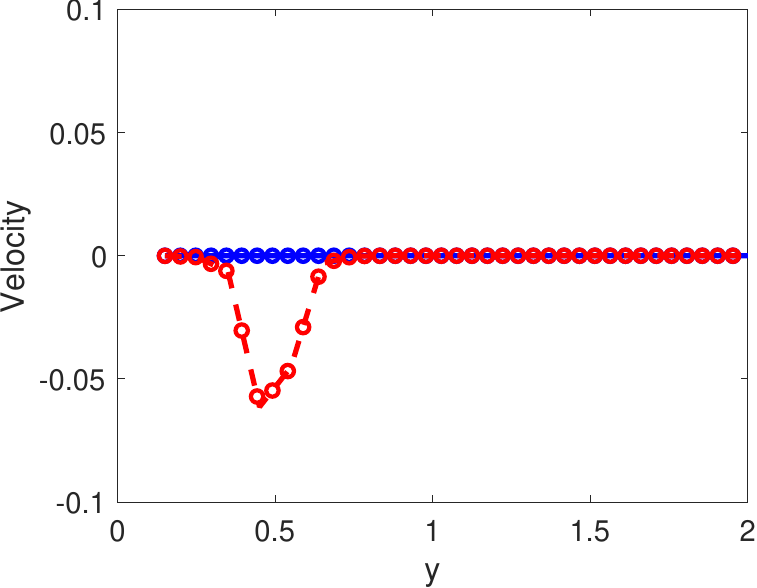}}
	\subfloat[]{\includegraphics[width=0.33\textwidth]{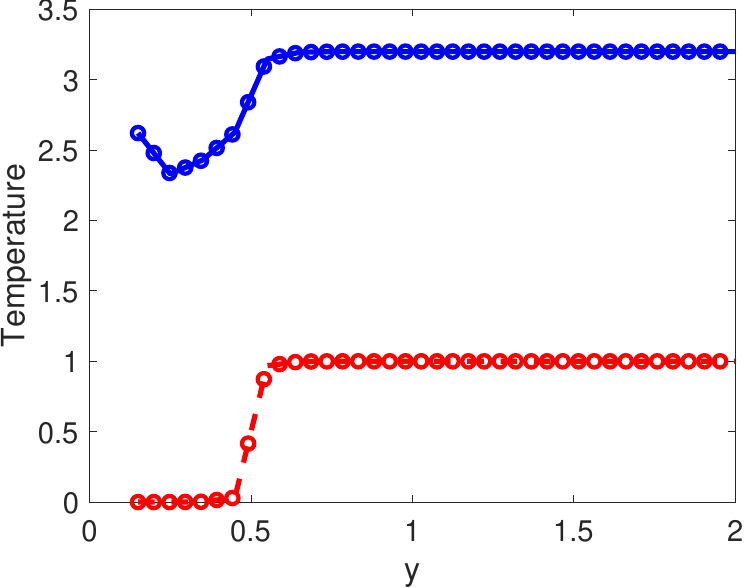}}
	\caption{Distributions of the electron and ion (a) density, (b) velocity, and (c) temperature with different mesh size at $x=2.0$. Lines: 20 mesh points in y-direction; Circles: 40 mesh points in y-direction.}
	\label{fig:mesh conver1}
\end{figure}

\begin{figure}[!htbp]
	\centering
	\subfloat[]{\includegraphics[width=0.33\textwidth]{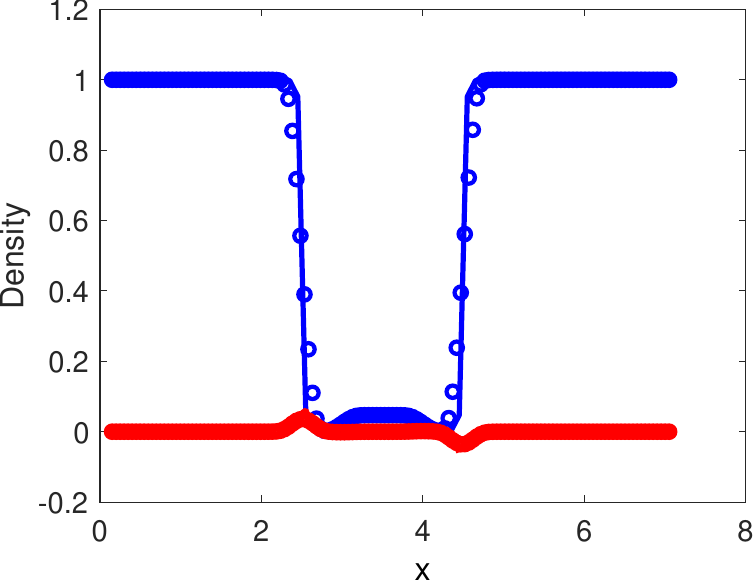}}
	\subfloat[]{\includegraphics[width=0.33\textwidth]{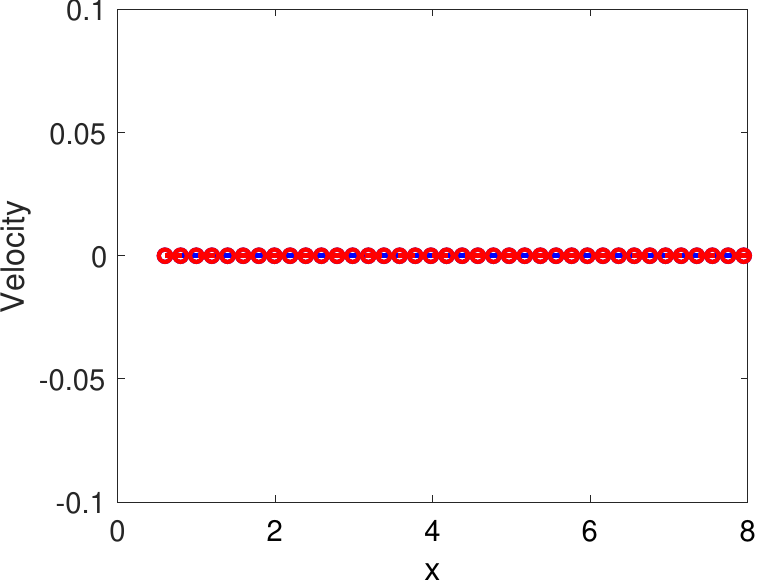}}
	\subfloat[]{\includegraphics[width=0.33\textwidth]{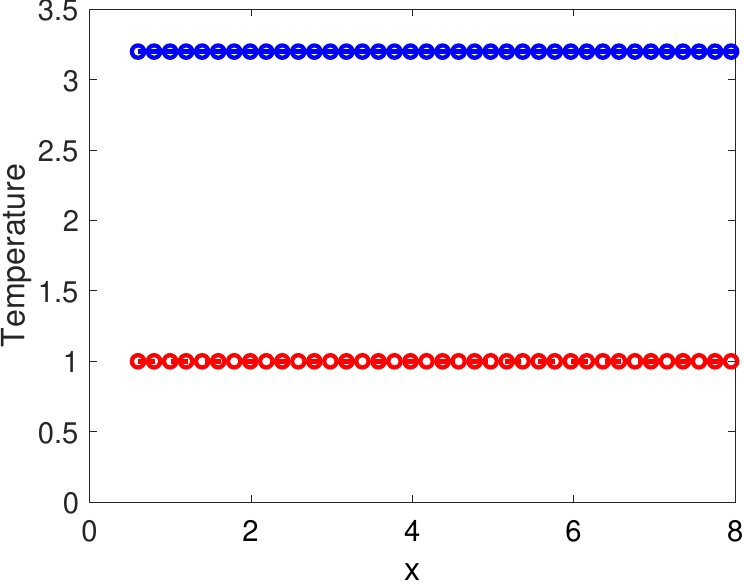}}
	\caption{Distributions of the electron and ion (a) density, (b) velocity, and (c) temperature with different mesh size at $y=1.0$. Lines: 70 mesh points in x-direction; Circles: 140 mesh points in x-direction.}
	\label{fig:mesh conver2}
\end{figure}

We also compare the $70\times20$ and $140\times40$ meshes to assess the spatial resolution of the heterogeneous flow. Both dimensions are refined by a factor of two. The material solver uses MUSCL reconstruction with a van Leer limiter and least-squares gradients. Figures~\ref{fig:mesh conver1} and~\ref{fig:mesh conver2} compare density, velocity, and temperature along $x=2.0$ and $y=1.0$ at $t=0.005$. The line profiles display the changes in the computed species fields under this refinement.

\FloatBarrier
\section{Conclusions}\label{sec:conclusion}
The coupled calculations show how species-dependent material transport and radiation exchange jointly shape radiation--plasma evolution. Binary-mixture shocks resolve distinct species profiles against Boltzmann reference results, and the Marshak wave connects the radiation update to diffusion behavior. The shock-tube and radiative-shock calculations exhibit coupled density, velocity, and thermal structures. In the heterogeneous Tophat problem, the paired simulations reveal changes in the spatial temperature response caused by radiation--material exchange and retain distinct electron, ion, and radiation fields.

These results establish the practical contribution of combining a multispecies material UGKS with angularly resolved radiation transport and velocity-dependent exchange terms. Relative to the earlier continuum-material formulation, the method evolves microscopic material distributions and couples their transport to radiative energy and momentum transfer. It provides a framework for investigating radiation--plasma flows where material collisions, photon interactions, and species relaxation vary across the domain.

\section*{Data Availability}

The data that support the findings of this study are available from the corresponding author upon reasonable request.




\bibliographystyle{siamplain}
\bibliography{ref}







\end{document}